\documentclass[aps,pre,reprint,showpacs,superscriptaddress]{revtex4-2}
\usepackage[utf8]{inputenc} 
\usepackage{tabularx}
\usepackage{chngcntr}
\usepackage{float}
\usepackage{color}
\usepackage{graphicx}
\usepackage{dcolumn}
\usepackage{bm}
\usepackage{mathrsfs}
\usepackage[ruled,vlined]{algorithm2e}
\usepackage{array} %
\usepackage{amssymb}
\usepackage[utf8]{inputenc}
\usepackage{tabstackengine}
\setstackEOL{\cr}
\usepackage{amsfonts,amsmath,mathrsfs,booktabs}
\usepackage{multirow,times}
\usepackage{float,color,bm}
\usepackage{empheq}
\usepackage{hhline}
\usepackage{tikz}
\usepackage{lipsum,mathtools}
\usepackage{diagbox}
\usepackage{bbm}
\usepackage[colorlinks,
linkcolor=blue,
anchorcolor=blue,
urlcolor=blue,
citecolor=blue
]{hyperref}
\newlength{\halfpagewidth}
\divide\halfpagewidth by 1

\newcommand{\splitatcommas}[1]{%
	\begingroup
	\ifnum\mathcode`,="8000
	\else
	\begingroup\lccode`~=`, \lowercase{\endgroup
		\edef~{\mathchar\the\mathcode`, \penalty0 \noexpand\hspace{0pt plus 0.3em}}%
	}\mathcode`,="8000
	\fi
	#1%
	\endgroup
}
\newcommand{\tuple}[1]{(\splitatcommas{#1})}
\newcommand{\set}[1]{\{\splitatcommas{#1}\}}
\newcommand{\RNum}[1]{\uppercase\expandafter{\romannumeral #1\relax}} 
\newcommand{\Rnum}[1]{\romannumeral #1\relax}
\newcommand{\PreserveBackslash}[1]{\let\temp=\\#1\let\\=\temp}
\newcolumntype{C}[1]{>{\PreserveBackslash\centering}p{#1}}
\newcolumntype{R}[1]{>{\PreserveBackslash\raggedleft}p{#1}}
\newcolumntype{L}[1]{>{\PreserveBackslash\raggedright}p{#1}}
 \normalsize

\allowdisplaybreaks[3]
\begin{document}
	\title{Dual Reinforcement Learning Synergy in Resource Allocation: Emergence of Momentum Strategy}
	\author{Zhen-Na Zhang}
	\address{School of Physics, Ningxia University, Yinchuan, 750021, P. R. China}
	\author{Guo-Zhong Zheng}
	\address{School of Physical Science and Technology, Inner Mongolia University, Hohhot 010021, P. R. China}
	\address{School of Physics and Information Technology, Shaanxi Normal University, Xi'an, 710062, P. R. China}
	\author{Li Chen}
	\address{School of Physics and Information Technology, Shaanxi Normal University, Xi'an, 710062, P. R. China}
	\author{Chao-Ran Cai}
	\address{School of Physics, Northwest University, Xi'an, 710127, P. R. China}
	\author{Sheng-Feng Deng}
	\address{School of Physics and Information Technology, Shaanxi Normal University, Xi'an, 710062, P. R. China}
	\author{Bin-Quan Li}
	\address{School of Physics, Ningxia University, Yinchuan, 750021, P. R. China}	
	\author{Ji-Qiang Zhang}\email{zhangjiqiang@nxu.edu.cn}\email{zhangjq13@lzu.edu.cn}
	\address{School of Physics, Ningxia University, Yinchuan, 750021, P. R. China}
	\date{\today}
\begin{abstract}
In ecosystems and human societies, resource allocation by self-organization and policy synergy is ubiquitous
and significant. This work focuses on the synergy between Dual Reinforcement Learning Policies in the
Minority Game (DRLP-MG) to optimize resource allocation. We examine a mixed-structured population
with two sub-populations: a subpopulation using Q-learning policy (Q-subpopulation) and the other adopting the
classical policy (C-subpopulation). We first identify a synergy effect between these subpopulations, where a first-order phase transition occurs as the mixing ratio of the two subpopulations changes. Further analysis reveals that the Q-subpopulation consists of two internal synergy clusters and one external synergy cluster. The former
contribute to the internal synergy through intra-subpopulation synchronization and anti-synchronization,
whereas the latter engages in the inter-subpopulation synergy. In the external synergy cluster, the well-known financial market momentum strategy emerges to play a key role in inter-subpopulation synergy and prevent long-term resource under-utilisation. However, the momentum strategy also triggers trend
reversals and leads to a decrease in rewards for those who adopt it. Notice that, our research reveals that the freezing effect  in either subpopulation is a crucial prerequisite for synergy, consistent with previous studies. We also conduct mathematical analyses on subpopulation synergy effects and the synchronization behaviors in the Q-subpopulation. Overall, our work systematically clarified the synergy mechanisms underpinning the complex resource-allocation dynamics of DRLP-MG, which offers valuable practical insights.

\end{abstract}
\maketitle
\section{Introduction}\label{introduction}
When confronted with the scarcity of resources, determining how to attain efficient resource allocation has emerged as a prominent and far-reaching concern in human society~\cite{ye2016survey,maritan2017resource,ibaraki1988resource,pitt2012axiomatization}. A key takeaway from classical economics is that self-organized markets can achieve this goal~\cite{smith1937wealth}. Not only that, the economics has also offered fundamental insights into the characteristics of optimal allocation from the general equilibrium theory~\cite{arrow1974general,debreu1982existence}. Nevertheless, the questions of how such an optimal state emerges via self-organization and under what evolutionary conditions remain unanswered. 

In the past decades, the rise of complexity science \cite{arthur2009complexity,epstein2014agent_zero} has given researchers a new way to explore macro-emergence via self-organized micro-interaction rules. It applies to studying collective behaviors like cooperation~\cite{nowak2012evolving,nowak2006five}, trust~\cite{engle2004evolution,yamagishi2011trust}, fairness~\cite{debove2016models,rand2013evolution}, and resource allocation~\cite{challet2004minority,challet2000relevance,marsili2001continuum,zhou2005self}.  A paradigmatic agent-based model for investigating resource allocation, which is derived from the El Farol bar problem~\cite{arthur1994inductive}, is the Minority Game (MG)~\cite{challet2004minority,challet2000relevance}.
In the MG, an odd number of agents repeatedly select between two resources with equal capacity to enter, and only those agents who choose the less-chosen (minority) resource emerge victorious while the others fail. Since the model was put forward, a vast number of studies have arisen regarding the fundamental collaborative mechanisms underlying the optimization of resource allocation~\cite{challet2000relevance,challet2004minority,michalopoulou2016employing}, particularly those associated with nonequilibrium phase transitions~\cite{challet2000relevance,galla2003dynamics,chakraborti2015statistical}. Moreover, inspired by the ``minority wins'' concept, some variants of the MG have been extended to network settings. In these variants, rather than relying on the rough-reinforcement-learning strategies used in the original MG, the strategies are pre-designed or are static rules that depend on local information~\cite{zhou2005self, paczuski2000self, dyer2008consensus, huang2012emergence,galstyan2002adaptive}. Within these variant models, researchers have carried out investigations on influence of pinning control on herding behavior~\cite{zhang2016controlling, zhang2013controlling}, the grouping phenomena
of resource selection in multi-resource systems~\cite{huang2012emergence} and so on~\cite{dyer2008consensus,paczuski2000self}.    

However, with the rapid development of reinforcement learning (RL)~\cite{sutton2018reinforcement,franccois2018introduction}, numerous studies have replaced the fixed strategy rules in conventional agent-based models with adaptive-feature-endowed RL to investigate emergence of human collective behaviors~\cite{zhang2020understanding,ding2023emergence,zheng2024decoding,zheng2023optimal,zheng2025decoding}.  
Consequently, some scholars also have attempted to use some newly developed RL strategies to substitute the original one for the investigation of the MG~\cite{zhang2019reinforcement,zhang2024self,shao2025network}. After using Q-learning in MG, they found it suppresses the resource under-utilization caused by the herding effect~\cite{andrecut2001q,zhang2019reinforcement}, similar to pinning control~\cite{zhang2016controlling, zhang2013controlling}. Also, a first-order phase transition distinct from classical MG was found in this case~\cite{zhang2024self}. These works provide a new perspective to further investigate the resource allocation in human society.  

Meanwhile, some findings have revealed that the diversity and heterogeneity inherent in strategies or interactions are capable of generating synergistic effects~\cite{geary2013understanding,kun2013resource,lee2023group,harmer1999parrondo}. Synergistic effects are widespread and play a crucial role in various fields. For example, the heterogeneity in resources \cite{kun2013resource,yan2023investigation}, group size~\cite{lee2023group}, and investment~\cite{yuan2014role,yan2023investigation} can jointly promote the emergence of cooperation. Such promotion is also found when multi-behavioral modes are incorporated~\cite{han2022hybrid,ma2023emergence,zhao2025evolution}, where the interaction of different modes synergistically reveals new complexities of cooperation. In epidemiology, different diseases or information may exhibit synergistic spread effects ~\cite{chen2019persistent,wang2019coevolution,liu2025coexistence}. In disease treatment, drugs or therapy strategies can act synergistically ~\cite{chen2015systematic, liu2025parrondo, li2024multifunctional,gu2012method,tarnopolsky2008mitochondrial}. As a result, an increasing number of studies are focusing on the synergy of different strategies~\cite{cheong2019paradoxical,liu2025parrondo}, especially the situation where combining losing strategies can yield winning results. In Ref.~\cite{cheong2019paradoxical}, the authors analyzed and found that the mechanism covers the entire biological spectrum. Moreover, other researchers have discovered that the synergy between different therapy strategies can effectively slow down the rapid development of tumors~\cite{liu2025parrondo}. 

In light of this line of thinking, we find ourselves especially captivated by the following query: \emph{Do synergies between reinforcement-learning-based policies also come into play in resource allocation?} Moreover, certain strategies offer advantages in resource utilization in reality, such as the momentum strategy of buying during market upswings and selling during downturns in financial markets~\cite{chan1996momentum}.
We then pose the following questions: \emph{Can one such strategy surface via reinforcement learning within a toy model? And under what conditions does this strategy emerge?} Answering these questions is essential for comprehending the synergy between policies in resource allocation within the framework of reinforcement learning.

The paper is structured as follows. In Section \ref{sec:model}, we introduce our Dual Reinforcement Learning Policies in the context of Minority Game (DRLP-MG), which is composed of two subpopulations. Specifically, one sub-population adopts the classical policy, while the other subpopulation employs the Q-learning policy. In Section \ref{sec:simulation}, we observe that there is an inter-subpopulation synergy in terms of resource utilization, and this synergy experiences a first-order phase transition. Based on the analysis in Section \ref{sec:analysis}, we find that the subpopulation using the Q-policy can be divided into two internal synergy clusters and an external synergy cluster. Moreover, the classical momentum strategy emerges in the external synergy cluster through the self-organization. Our conclusions and discussions are presented in Sec.~\ref{sec:summary}.

\section{Model}\label{sec:model}
In this study, we initially present our DRLP-MG model. In the model, the population consists of an odd-number of agents, denoted by a set $\mathcal{N}$. 
At each Monte Carlo step $\tau$ in the evolutionary dynamics, the protocol involves two processes: \emph{gaming} and \emph{learning} processes. 
During the gaming process, each agent $i\in\mathcal{N}$ will take an action $a^i$ from the action set $\mathcal{A} = \set{0, 1}$ according to its policy as the entering resource. Here, it is noted that the resource set $\mathcal{R}$ is the same as $\mathcal{A}$. 

For each resource $r\in\mathcal{R}$, it has a capacity for agents that is $C_{r}  = \lfloor{\frac{N}{2}}\rfloor$. If the number of agents entering the resource $r$ at $\tau$ is below its capacity $C_{r}$, then the resource $r$ is the winning resource $r_w$ at the step, i.e.,
\begin{equation}
	r_{w}(\tau) = 
	\begin{cases}
		1 & \text{if}\sum\limits_{i\in{\mathcal{N}}}a^{i}(\tau)\le C_{1}, \\
		0 & \text{otherwise}. 
	\end{cases}
	\label{eq:winning_resource}
\end{equation}
For any agent $i\in\mathcal{N}$, if it enters the resource $r_{w}$ at the step it will receive a reward of $1$; otherwise $-1$, i.e.,
\begin{eqnarray}
	\Pi^{i}(\tau) = \mathbbm{1}_{a^{i}(\tau) = r_{w}(\tau)}
	- \mathbbm{1}_{a^{i}(\tau) \ne r_{w}(\tau)}.
	\label{eq:payoff}
\end{eqnarray} 
Here, $\mathbbm{1}_{\emph{predicate}}$ denotes the variable that is $1$ if $predicate$ is true and $0$ if it is not, 

Two reinforcement learning policies are available for the agents in the population: the classical policy $\pi_c$ and the Q-learning policy $\pi_q$ designed by us. Under both policies, agents take actions based on the common information, which is the historical winning resources over the past $m$ steps. This is called a state and is denoted as $s(\tau) = r_w(\tau-m)r_w(\tau-m+1)\cdots r_w(\tau-1)$ at $\tau$th step. Then, $s(\tau)$ can be further converted into a decimal number $s_{\mu}$ in $[0, 2^{m}-1]$ to label,
\begin{equation}
	s_{\mu} = \sum_{k = 1}^{m}2^{k-1}\cdot r_{w}(\tau-k).
	\label{eq:state}
\end{equation}
Thus, the state set can be represented as $\mathcal{S} = \set{s_{0}, s_{1}, \cdots, s_{2^{m}-1}}$.  

In our model,  the entire population $\mathcal{N}$ is composed of two structured mixed subpopulations.  Specifically, these are C-subpopulation $\mathcal{N}_{c}$ and Q-subpopulation $\mathcal{N}_{q}$, and it holds $\mathcal{N} = \mathcal{N}_{c} \cup \mathcal{N}_{q}$. The agents in $\mathcal{N}_{c}$ employ the classical policy $\pi_{c}$~\cite{challet2004minority,moro2004minority}, while the agents in $\mathcal{N}_{q}$ make use of the Q-learning policy $\pi_{q}$. The detail for these policies is as follows: 

{\bf Classical Policy}-- In the game process of $\pi_{c}$, any agent $i$ takes action based on the current state and strategies in its strategy base $\mathcal{B}^{i}$. In the base, each strategy $\hat{\bm{a}}\in\mathcal{B}^{i}$ is a state-action map. This map is formed by a binary Bernoulli sequence with a length of
$2^m$ and is denoted as $\tuple{\hat{a}_0, \hat{a}_1, \cdots, \hat{a}_{2^m-1}}$ [see Table.~\ref{tab:policies}]. 
And, $\hat{a}_\nu$ in $\hat{\bm{a}}$
represents the action that the agent takes in state $s_\nu$ when it adopts $\hat{\bm{a}}$.
In the game process of $\tau$th step, $i$ selects the strategy with the highest accumulated score from $\mathcal{B}^i$, and then takes the corresponding action for the current state $s_{\mu}$ according to this strategy, i.e.,
\begin{subequations}	
 	\begin{empheq}[left=\empheqlbrace]{align}
	\hat{\bm{a}}^{i}(\tau) &= \arg \max_{\hat{\bm{a}}^{\prime}\in \mathcal{B}^{i}}\text{score}(\hat{\bm{a}}^{\prime}, \tau), \\
	a^{i}(\tau)  &= \pi_c(\hat{\bm{a}}^{i}(\tau), s_{\mu}) = \hat{a}^{i}_{\mu}. 
	\end{empheq}\label{eq:classical_action}
\end{subequations}
In Eq.~\eqref{eq:classical_action}, $\hat{\bm{a}}^{i}$ and $a^{i}$ are the currently selected strategy and action for agent $i$,  respectively. And, $\hat{a}^{i}_{\mu}$ denote the $\mu$th element in $\hat{\bm{a}}^{i}$. Here, it should be noted that if there is more than one strategy with the highest cumulative score, then $i$ randomly selects one of them as its current strategy $\hat{\bm{a}}^{i}$. Evidently, within each strategy $\hat{\bm{a}}$, the mapping actions for different states are integrated rather than independent.

In the learning process, the new winning resource $r_{w}(\tau)$ can be gained as per Eq.~\eqref{eq:winning_resource} once all agents have taken actions. According to $r_{w}(\tau)$,    
$i$ will update the score of each strategy $\hat{\bm{a}}\in\mathcal{B}^{i}$ based on the real or virtual gain achieved.
The detail is as follows:
\begin{equation}
\text{score}(\hat{\bm{a}}, \tau) = \text{score}(\hat{\bm{a}}, \tau - 1) + \mathbbm{1}_{\hat{a}_{\mu} = r_{w}(\tau)}
- \mathbbm{1}_{\hat{a}_{\mu} \ne r_{w}(\tau)},
\label{eq:update_score}
\end{equation}
in which $\hat{a}_{\mu}$ is the $\mu$th element in $\hat{\bm{a}}$.

\begin{table}[htbp!]
	\begin{tabular}{|p{0.1cm}|p{0.1cm}p{0.1cm}p{0.1cm}|p{0.1cm}p{0.1cm}|cc|}
		\hline
		\multicolumn{4}{|c|}{State} & \multicolumn{2}{c|}{Base $\mathcal{B}$} & \multicolumn{2}{c|}{Q-table} \\
		\hline
		\quad&\multicolumn{3}{c|}{History} & $\hat{\bm{a}}_1$ & $\hat{\bm{a}}_2$ &  0 & 1\\
		\hline
		$s_0$ &  0 & 0  & 0 &  1 &  0 &  $Q_{s_0,0}$ & $Q_{s_0,1}$ \\
		$s_1$ &  0 & 0  & 1 &  1 &  1 &  $Q_{s_1,0}$ & $Q_{s_1,1}$ \\
		$s_2$ &  0 & 1  & 0 &  0 &  1 &  $Q_{s_2,0}$ & $Q_{s_2,1}$ \\ 
		\vdots & \vdots & \vdots & \vdots&  \vdots& \vdots &  \vdots & \vdots\\
		$s_7$ &  1 & 1  & 1 &  0 &  0 &  $Q_{s_7,0}$ & $Q_{s_7,1}$ \\
		\hline
	\end{tabular}
	\caption{{\bf An instance of the classical policy on the strategy base and of the Q-learning policy on the Q-table.} In the first three columns of the table, the instance lists all possible states, which represent the historical winning resources over the past $m = 3$ steps.
	In the subsequent columns, as an example, a strategy base 
	 $\mathcal{B}$ that includes $|\mathcal{B}| = 2$ strategies, $\hat{\bm{a}}_1$ and $\hat{\bm{a}}_2$, 
	for the classical policy $\pi_{c}$ is presented. The last columns display the state-action values of all Cartesian products of states and actions for the Q-learning policy $\pi_{q}$.
	 }\label{tab:policies}
\end{table}

{\bf Q-learning policy}-- In the game process of $\pi_{q}$, agents execute actions according to the cognitive action values of different actions in the current historical state, following the Q-learning algorithm~\cite{sutton2018reinforcement}. Without loss of generality, we will introduce the algorithm by taking a specific agent $i$ as an example. For the agent $i$, the cognitive action values of different actions in different states are represented as a mapping from the Cartesian product of states (columns) and actions (rows), $\mathcal{S}\times\mathcal{A}\rightarrow \mathbb{R}$, and they form a Q-table. At $\tau$th step, $i$ selects the corresponding action with the maximum cognitive action value in the current state $s_{\mu}$ 
with probability $1-\epsilon$, or a random action within $\mathcal{A}$ otherwise. The equivalent detail is as follows:
\begin{align}
	a^{i}(\tau) &= \pi_{q}(s_\mu, {\bf Q}^{i}(\tau)) \nonumber \\
	&= \left\{
	\begin{aligned}
		&\arg\max\limits_{a'}\left\{Q_{s_\mu,a}(\tau)\right\}, &  1 -\varepsilon + \frac{\varepsilon}{|\mathcal{A}|};  \\
		&1 - \arg\max\limits_{a'}\left\{Q_{s_\mu,a}(\tau)\right\},  & \frac{\varepsilon}{|\mathcal{A}|}.
	\end{aligned}	
	\right.
	\label{eq:reinforce_action}
\end{align}
Here, $\varepsilon\in[0, 1]$ is a parameter that determines the trade-off between exploitation and exploration. At the end of game process, $i$ will receive its reward $\Pi^{i}$ as mentioned before.

In the learning process, $i$ update the element $Q_{s_{\mu},a^{i}}$ for its Q-table as follows: 
\begin{equation}
   Q_{s_{\mu},a^{i}}(\tau + 1) = (1-\alpha)Q_{s_{\mu},a^{i}}(\tau) +\alpha\left(\gamma Q^{\max}_{s',a'}(\tau) + \Pi^{i}(\tau)\right).\label{eq:Q_update}
\end{equation}  
Here, $\alpha \in (0, 1]$ is the learning rate reflecting the influence of new experience on the old. $\gamma\in [0, 1)$ is the discount factor determining the importance of future rewards since $ Q^{\max}_{s',a'}$ is the maximum action value in the row of next state $s^{\prime} = s(\tau + 1) = r_{w}(\tau-m+1)r_{w}(\tau-m+2)\cdots r_{w}(\tau)$ that could be expected.
\begin{algorithm}[htbp!]  
	\caption{Algorithm for DRLP-MG \\in the mixed population}\label{algorithm:protocol}
	\LinesNumbered 
	\KwIn{Learning parameters: $\alpha$, $\gamma$, $\varepsilon$; Memory length: $m$; Size of strategy base: $|\mathcal{B}|$;
	Population: $\mathcal{N}$; Subpopulations: $\mathcal{N}_{c}$ and $\mathcal{N}_{q}$}
	{\bf Initialization}\;
	{Create a randomly-generated history $\bm{s}$ with a length of $m$}\;
	\For{$i$ in $\mathcal{N}_c$}{
		Create a strategy base $\mathcal{B}^{i}$ include $|\mathcal{B}^{i}|$ strategies\;
		\For{$\hat{\bm{a}}$ in $\mathcal{B}^i$}{ 
			Initialize the score of the strategy $a$ to zero\; 
		}	
	}
	\For{$i$ in $\mathcal{N}_q$}{
		Create a Q-table with each item in the matrix near zero\;	
	}
	\Repeat{\text the system becomes statistically stable or evolves for the desired time duration}
	{{\bf Gaming process}\;
		\For{$i$ in $\mathcal{N}_c$}{
			Take action $a^i$ according to state, scores and Eq.~\eqref{eq:classical_action}\;
		}
		\For{$i$ in $\mathcal{N}_q$}{
			Generate a random number $p$\;
			\eIf{$p<\varepsilon$}{
				Pick an action randomly from $\mathcal{A}$
			}{
				Take action $a^{i}$ according to the current state, Q-table and Eq.~\eqref{eq:reinforce_action}
			}
		}
		{Get the winning resource $r_{w}$ according to \eqref{eq:winning_resource}\;
			\For{$i$ in $\mathcal{N}$}{
				Get reward $\Pi^{i}$ according to Eq.~\eqref{eq:payoff}\; 
			}
			{{\bf Learning process}\;
				{Get next state $s^{\prime}$}}\;
			\For{$i$ in $\mathcal{N}_c$}{
				\For{$\hat{\bm{a}}\in\mathcal{B}^i$}{
					Update the score of $\hat{\bm{a}}$ according to Eq.~\eqref{eq:update_score}\;
				}
			}			
			\For{$i$ in $\mathcal{N}_q$}{	
				Update Q-table according to Eq.~(\ref{eq:Q_update})\;
			}
		}
		{Update state $\bm{s}$ as $\bm{s}^{\prime}$}
	}
\end{algorithm}

At the end of each step, the new winning resource $r_{w}(\tau)$ can be gained based on Eq.~\eqref{eq:winning_resource}. Then, the state is updated to $s^{\prime}$ according to new historical winning resources over the past $m$ steps. Two obvious differences between $\pi_{c}$ and $\pi_{q}$ are as follows: 1) The learning granularity for $\pi_{c}$ is notably coarser than that for $\pi_{q}$ because $\pi_{c}$ focuses on the optimal strategy rather than directly on the optimal state-action relation as in $\pi_{q}$; 2) The virtual scores within $\pi_{c}$ may diverge, whereas the state-action values of $\pi_{q}$ are guaranteed to converge. In the simulation, the evolving processes are repeated until the system reaches statistical stability or the desired time duration has elapsed. The pseudo-code of our model is presented in Algorithm.~\ref{algorithm:protocol}.

For our model, the optimal resource allocation is $\sum_{i\in\mathcal{N}}a^{i}(\tau) = C_1$, i.e., the number of agents entering the resource is equal to its capacity. To measure the performance of the population, we employ the \emph{volatility} of the capacity $C_{1}$,
\begin{equation}
	\psi := \frac{\sigma^2}{|\mathcal{N}|} = \frac{\sum\limits_{\tau = t_0}^{T}(N_1(\tau) - C_1)^2}{|\mathcal{N}|(T-t_0)},
	\label{eq:volatility}
\end{equation}
to characterize the statistical deviation from the optimal resource utilization over $T-t_0$ steps~\cite{challet2004minority,moro2004minority,zhang2016controlling}. 
Here, $t_0$ represents an arbitrary step at which the system has reached statistical stability. And, $N_1(\tau) = \sum_{i\in\mathcal{N}}a^i(\tau)$ is the number of agents entering resource $1$. Evidently, a lower $\psi$ indicates a higher efficiency resource allocation.
	
Furthermore, it is crucial to examine the volatility of resource selection within the C-subpopulation $\mathcal{N}_{c}$ and the Q-subpopulation $\mathcal{N}_{q}$ that are
\begin{equation}
	\psi_{c(q)} := \frac{\sigma^2_{c(q)}}{|\mathcal{N}_{c(q)}|} = \frac{\sum\limits_{\tau = t_0}^{T}\left(N_{c_1(q_1)}(\tau) - \overline{N}_{c_1(q_1)}\right)^2}{|\mathcal{N}_{c(q)}|(T-t_0)}.
	\label{eq:sub_volatilities}
\end{equation}
Here, $N_{c_1(q_1)}(\tau) = \sum_{i\in\mathcal{N}_{c(q)}}a^i(\tau)$ is the number of agents within $\mathcal{N}_{c(q)}$ who enter resource $1$. And $\overline{N}_{c_1(q_1)}$ is the average of $N_{c_1(q_1)}$ over the time interval $T-t_0$.  For simplicity, $\psi_{c}$ and $\psi_q$ are referred to as \emph{C-volatility} and \emph{Q-volatility}, respectively.

In this study, our primary objective is to comprehend how the inter-subpopulation synergy changes in relation to $f_{c}$, where $f_{c}$ represents the fraction of the C-subpopulation within the entire population. In addition, the fractions  $f_{c}=|\mathcal{N}_{c}|/|\mathcal{N}|$ and $f_{q}=|\mathcal{N}_{q}|/|\mathcal{N}|$ meet $f_{c} + f_{q} = 1$. Without specific declaration, the default learning parameters are set as $(\alpha, \gamma, \varepsilon) = (0.1, 0.9, 0.01)$, memory length is $m = 3$, the  size of the strategy base is $|\mathcal{B}|=2$ and the size of system is $|\mathcal{N}| = 301$.

\section{Simulation Results}\label{sec:simulation}
\begin{figure}[htbp!]
	\centering
	\includegraphics[width=0.9\linewidth]{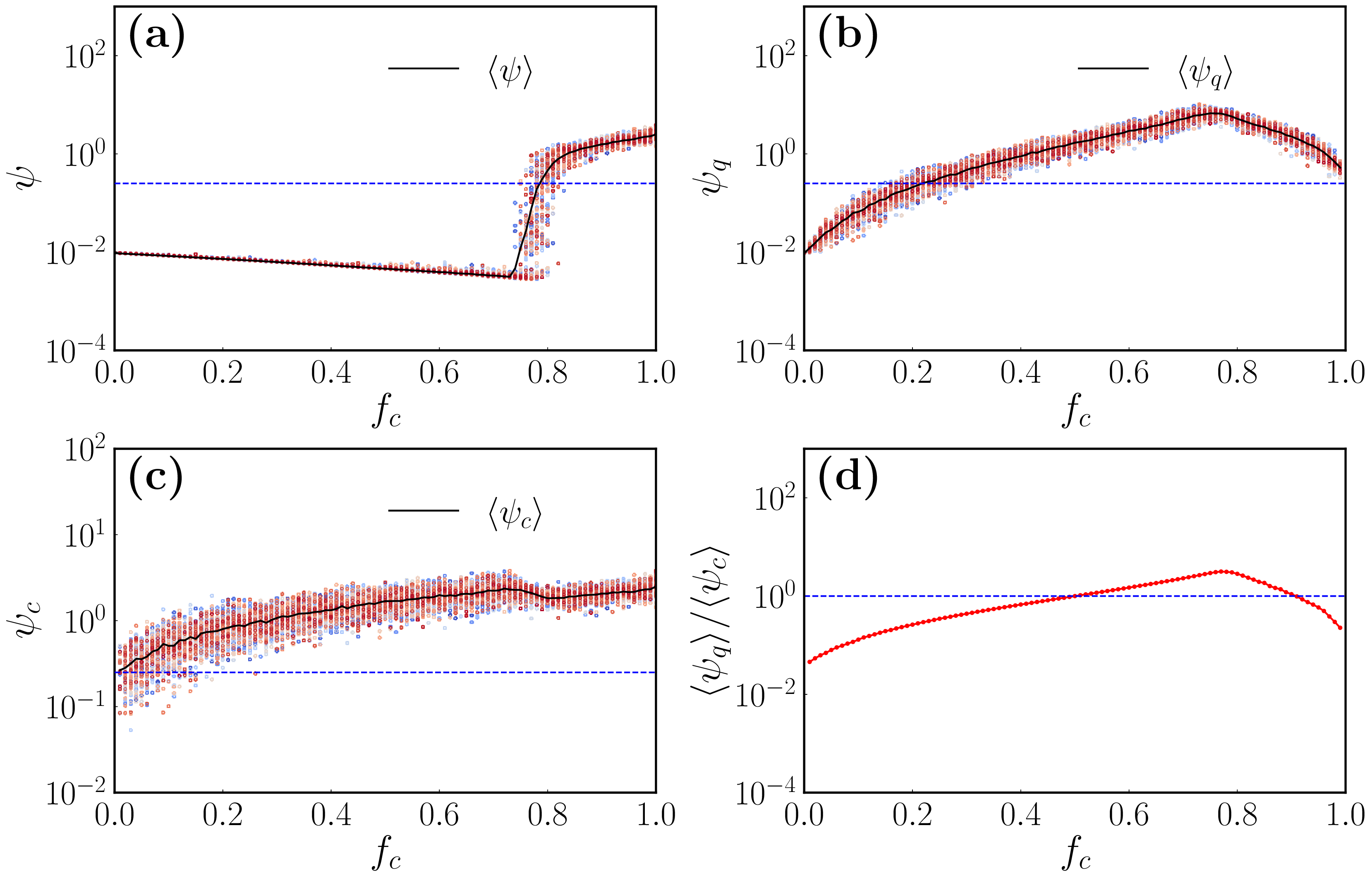}
	\caption{ (Color online) {\bf Volatility within the whole population and subpopulations, along with the ratio of volatility between the two subpopulations.} (a) illustrates the volatility $\psi$ within the whole population as a function of $f_{c}$. (b) and (c) respectively illustrate the C-volatility $\psi_{c}$ and Q-volatility $\psi_{q}$ as functions of $f_{c}$. For each $f_{c}$, the outcome of each individual run within the ensemble is denoted by dots of a specific color. Meanwhile, the ensemble average  $\langle\cdots\rangle$ calculated over $100$ runs is represented by a black line. In panels (a) to (c), the blue dotted line marks $\psi = 0.25$, $\psi_c = 0.25$ and $\psi_q = 0.25$ of the random-choice system respectively, each serving as its corresponding benchmark. In (d), the plot shows the ratio $\langle\psi_{q}\rangle/\langle\psi_{c}\rangle$ as the function of $f_{c}$ and $\langle\psi_{q}\rangle/\langle\psi_{c}\rangle = 1$ marked with a blue dotted line.   
	The default learning parameters are $(\alpha, \gamma, \epsilon) = (0.1, 0.9, 0.01)$, memory length is $m = 3$, the  size of the strategy base is $|\mathcal{B}|=2$, and system size is $|\mathcal{N}| = 301$. 
	}\label{fig:volatility}
\end{figure}

Figure \ref{fig:volatility}(a - c) first shows how the fraction $f_{c}$ affects the volatility $\psi$ of the capacity, as well as the sub-volatility $\psi_{c}$ and $\psi_{q}$ within subpopulations. In Fig.~\ref{fig:volatility}(a), one learns that  $\psi$ is non-monotonic with the increase of $f_{c}$. Before the transition point $f_{c}^{*}$, $\psi$ exponentially decreases with the increase of $f_{c}$, after this point, it suddenly increases and keeps a slow increase eventually as $f_{c}$ rises. Furthermore, upon observation, even when the entire population is made up of only Q-subpopulations, the resource allocation achieved through self-organization outperforms that under the random choice game. This suggests the possible existence of an \emph{intra-synergy} within the Q-subpopulation. Moreover, $\psi$ in different runs of the ensemble exhibit significant fluctuations around the point $f_{c}^{*}$, suggesting that a phase transition takes place at $f_{c}^{*}$. 

Different from volatility in the whole population, the Q-volatility $\psi_{q}$ starts by increasing. Then, after experiencing a slight decline in the vicinity of $f_{c}^{*}$, it begins to increase once more [see Fig.~\ref{fig:volatility}(b)]. Meanwhile, the C-volatility $\psi_{c}$ also first increases when $f_{c}$ is less than the transition point $f_{c}^{*}$, and then decreases as $f_{c}$ continues to increase beyond $f_{c}^{*}$[see Fig.~\ref{fig:volatility}(c)]. Based on Fig.~\ref{fig:volatility} (a-c), a remarkable phenomenon is observed: $\psi$, the volatility of the entire population consistently remains lower than both $\psi_c$ and  $\psi_q$ within the subpopulations. The findings suggest that beyond the intra-synergy existing within the Q-subpopulation, inter-subpopulation synergy also takes place. This inter-synergy results in a more in-depth optimization of resource allocation in the case that $f_{c}\le f_{c}^{*}$.

In Fig.~\ref{fig:volatility}(d), the influence of $f_{c}$ on the gap between the C-volatility and Q-volatility is depicted through the ratio $\langle\psi_{q}\rangle/\langle\psi_{c}\rangle$. Similar to change of $\psi_{q}$ as $f_{c}$
increases, the ratio $\langle\psi_{c}\rangle/\langle\psi_{q}\rangle$ also initially rises when $f_{c}<f_{c}^*$ and 
subsequently falls when $f_{c}>f_{c}^*$. Moreover, there is an interval within which $\langle\psi_{c}\rangle/\langle\psi_{q}\rangle$ increases exponentially with $f_{c}$. Additionally, in this particular context, the gap between $\langle\psi_{c}\rangle$ and $\langle\psi_{q}\rangle$ disappears when $f_{c} \approx 0.5$, i.e., the sizes of the two subpopulations are equal.

\begin{figure}[htbp!]
	\centering
	\includegraphics[width=0.9\linewidth]{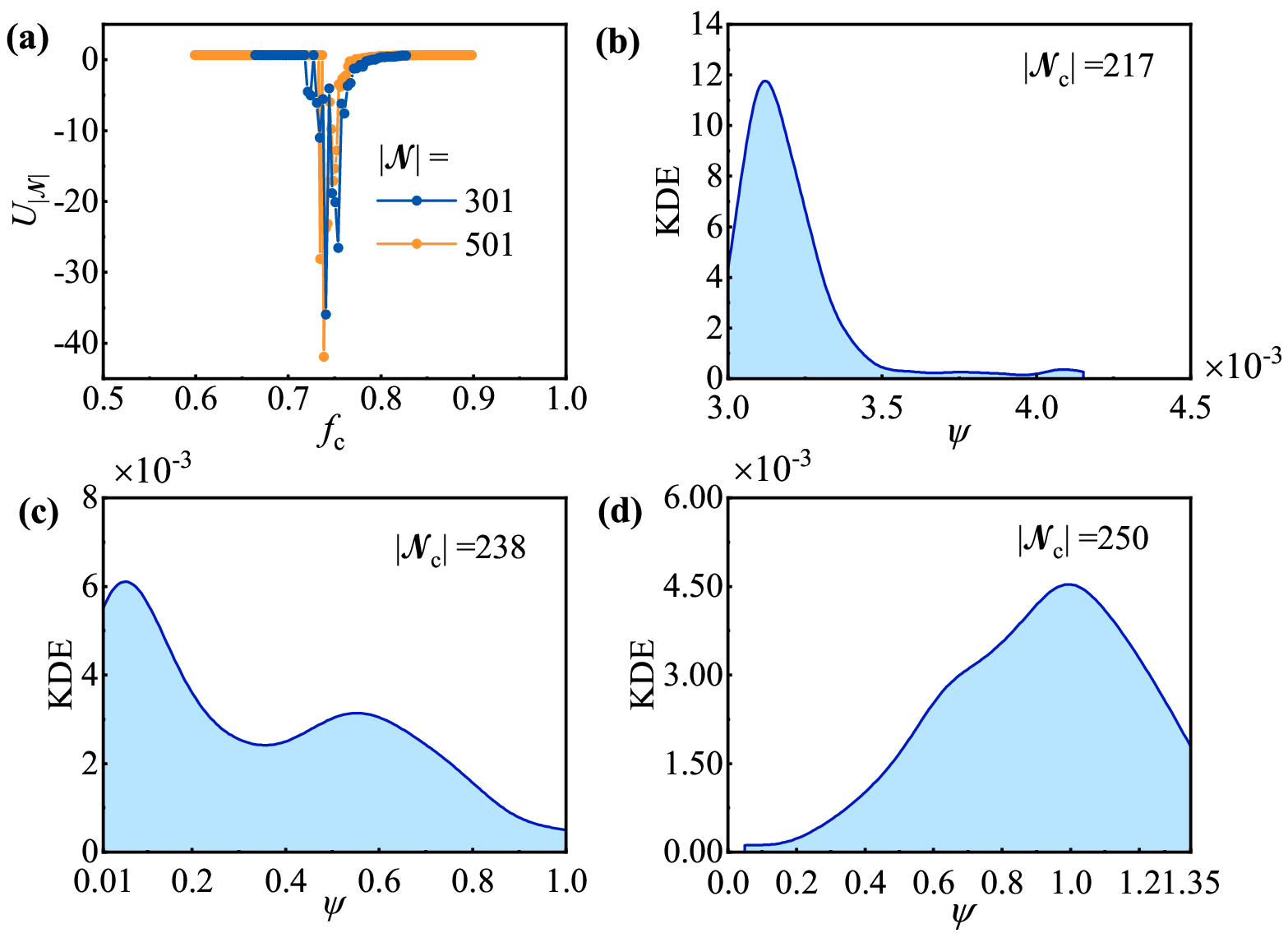}
	\caption{ (Color online) {\bf The Binder cumulant of volatility as the function of the fraction of the classical agents and the Kernel Density Estimation (KDE) of volatility of the ensemble.} Panel (a) depicts the relationship between the Binder cumulant of volatility $\psi$ and $f_c$ in an ensemble consisting of $300$ runs. (b - d) show the KDEs of $\psi$ at different values of $f_c$ for the ensemble, facilitating a detailed examination of the probability distribution across different runs. In (a–d), the parameters remain at their default settings, with only an additional result under another system size $|\mathcal{N}| = 501$ being added in (a).
	}\label{fig:Binder_kde}
\end{figure}

To further determine the type of phase transition of $\psi$ at 
$f^*_{c}$, we first illustrate the relationship between the Binder cumulant of volatility and $f_{c}$, where the Binder cumulant~\cite{binder1981finite} is defined as
\begin{equation}
	U_{|\mathcal{N}|} := 1 - \frac{\langle\psi^4\rangle_{|\mathcal{N}|}}{3\langle\psi^2\rangle_{|\mathcal{N}|}^2}.
	\label{eq:binder}
\end{equation}
As shown in Fig.~\ref{fig:Binder_kde} (a), the curves of $U_{|\mathcal{N}|}(f_c)$ for systems with different sizes do not intersect at one common point. A distinct inverted peak appears in the Binder cumulant around 
$f_c^{*}\approx 0.73$ and its height increases as the system size $|\mathcal{N}|$ increase. The result indicates that a first-order phase transition occurs at $f_{c}^{*}$. 

To further verify the indication, we then show the Kernel Density Estimation (KDE)~\cite{davis2011remarks} of $\psi$ of an ensemble at different $f_{c}$ in Fig.~\ref{fig:Binder_kde} (b-d). (b) shows that when $f_{c} < f_{c}^{*}$, KDE exhibits only one peak at a low-volatility.  In contrast, as shown in Fig.~\ref{fig:Binder_kde} (c), when
$f_{c}$ is around $f_{c}^{*}$, another peak emerges at a high-volatility, coexisting with the previous one. When $f_{c} > f_{c}^{*}$, the peak corresponding to low volatility vanishes, whereas the high-volatility peak persists [see Fig.~\ref{fig:Binder_kde} (d)]. The results further demonstrate that as the parameter 
$f_{c}$ increases, the volatility $\psi$, serving as an order parameter, experiences a first-order phase transition near the phase transition point. Specifically, the low-volatility gradually loses its stability and is replaced by the high-volatility whose stability gradually increases.

\section{Mechanism Analysis}\label{sec:analysis}
\subsection{Inter-subpopulation synergy}
\begin{figure}[htbp!]
	\centering
	\includegraphics[width=0.95\linewidth]{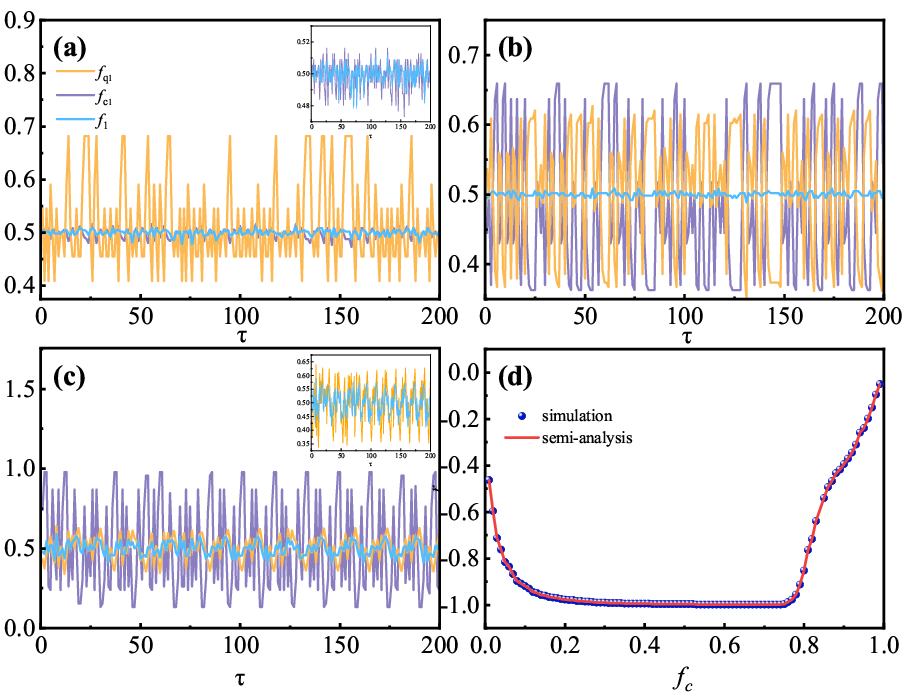}
	\caption{ (Color online) {\bf Time series of agents’ fractions entering a specific resource in subpopulations and entire population, plus Pearson correlation coefficient of subpopulation entry-fraction time series.} (a-c) show the time series of $f_{c_1}$, $f_{q_1}$ and $f_{1}$ under different fraction of C-subpopulation, where the values of these fractions are $f_{c} = 0.075$, $0.45$, $0.85$, respectively. The insets of (a) and (c) are the local zoom-ins of the time series  presented in the main panel.
(d) exhibits the Pearson coefficient between the time series of  $N_{c_1}$ and $N_{q_1}$ as a function of $f_{c}$. This is presented based on Eq.~\eqref{eq:pearson_corraltion} in the simulation and Eq.~\eqref{eq:psi_psicq} in the semi-analytical approach. The learning parameters $(\alpha, \gamma, \epsilon)$, memory length $m$, size of strategy base $|\mathcal{B}|$ and system size $|\mathcal{N}|$ are set as those in Fig.~\ref{fig:volatility} by default. 
	}\label{fig:timeseries_Pearson}
\end{figure}

To further demonstrate the inter-subpopulation synergy in terms of resource allocation as implied by Fig.~\ref{fig:volatility} (a - c), we present the time series of the fractions of agents entering resource $1$ in $\mathcal{N}_c$ and $\mathcal{N}_q$, as well as in the entire population $\mathcal{N}$ under different $f_{c}$ [see Fig.~\ref{fig:timeseries_Pearson} (a-c)]. The definitions for these fractions are 
\begin{subequations}\label{eq:f1_fc1_fq1}
\begin{empheq}[left=\empheqlbrace]{align}
			&f_{c_1(q_1)}(\tau) := \displaystyle{\frac{N_{c_1(q_1)}(\tau)}{|\mathcal{N}_{c(q)}|}} = \frac{\sum\limits_{i\in\mathcal{N}_{c(q)}}a^i(\tau)}{|\mathcal{N}_{c(q)}|},\\
			&f_{1}(\tau) := \displaystyle{\frac{N_{1}(\tau)}{|\mathcal{N}|}} = \displaystyle{\frac{\sum\limits_{i\in\mathcal{N}}a^i(\tau)}{|\mathcal{N}|}},
 \end{empheq}
\end{subequations}
which meet $f_1(\tau) = f_{c}f_{c_1}(\tau) + f_{q}f_{q_1}(\tau)$. The results show $f_{c_1}$, $f_{q_1}$ and $f_{1}$ all oscillate around their respective means $\bar{f}_{c_1}$, $\bar{f}_{q_1}$ and $\bar{f}_{1}$. However, the fluctuations of $f_{c_1}$ 
and $f_{q_1}$ exhibit a negative correlation and are both smaller than the fluctuation of $f_{1}$. In addition, the fluctuations of $f_{c_1}$ and $f_{q_1}$ are negatively associated with $f_{c}$ and $f_{q}$, respectively.
The results indicate that optimizing resource allocation across the entire population can be achieved through the strong inter-synergy, rather than merely the simple combination of individual optimizations within each subpopulation. In addition, the inter-synergy effect is manifested in the negative correlation between $f_{c_1}$ and $f_{q_1}$.

Moreover, Fig.~\ref{fig:timeseries_Pearson} (a-c) further display $\bar{f}_{1}$ always approach to the optimal allocation $1/2$ for different $f_{c}$. However, both $\bar{f}_{c_1}$ and $\bar{f}_{q_1}$ deviate from $1/2$ and these deviations are denoted as $\Delta\bar{f}_{c_1}$ and $\Delta\bar{f}_{q_1}$, respectively. By carefully examination, we learn that $\Delta\bar{f}_{c_1}$ and $\Delta\bar{f}_{q_1}$ are also negatively related to $f_{c}$ and $f_{q}$. This pattern is consistent with the fluctuations of $f_{c_1}$ and $f_{q_1}$. The analysis in Appendix~\ref{sec:app_volatility} further reveals that these deviations satisfy
\begin{align}\label{eq:deviation}
	f_{c}\Delta\bar{f}_{c_1}+f_{q}\Delta\bar{f}_{q_1} \approx 0, 
\end{align}
if the resources are well-allocated for the population. Finally, similar to the findings in some previous studies~\cite{zheng2023optimal}, the results show that both $f_{c_1}$ and $f_{q_1}$ are confined to certain discrete values within a specific region, rather than continuously covering the entire region.

To conduct a further investigation into the inter-synergy, we present the Pearson coefficient between the time series of 
$N_{c_1}$ and $N_{q_1}$
as a function of $f_{c}$ in Fig.~\ref{fig:timeseries_Pearson} (d). Below is the definition of the Pearson correlation coefficient
\begin{align}
	r := \frac{\sum\limits_{\tau=t_0}^{T}(N_{c_1}(\tau)-\overline{N}_{c_1})(N_{q_1}(\tau)-\overline{N}_{q_1})}
	{\sqrt{{\sum\limits_{\tau = t_0}^{T}(N_{c_1}(\tau)-\overline{N}_{c_1})^2}}{\sqrt{{\sum\limits_{\tau = t_0}^{T}(N_{q_1}(\tau)-\overline{N}_{q_1})^2}}}}.
	\label{eq:pearson_corraltion}
\end{align}
where $\overline{N}_{c_1(q_1)}$ is the average of $N_{c_1(q_1)}$ over $T-t_0$. In Appendix~\ref{sec:app_volatility}, we derive the relation between C-volatility $\psi_c$, 
Q-volatility $\psi_{q}$ and overall volatility $\psi$, which is given by 
\begin{align}\label{eq:psi_psicq}
	\psi &=  f_{c}\psi_{c}+f_{q}\psi_{q} + 2r\sqrt{f_{c}\psi_{c}}\sqrt{f_q\psi_{q}},
\end{align}
and presented in Fig.~\ref{fig:timeseries_Pearson} (d).
Clearly, a negative correlation between $\psi_{c}$ and $\psi_{q}$ reduces $\psi$, while a positive one amplifies it. 
\subsection{Analysis of Q-subpopulation}
\subsubsection{Internal and external synergy clusters}\label{subsubsec:Intra_synergy}
\begin{figure}[htbp!]
	\centering
	\includegraphics[width=\linewidth]{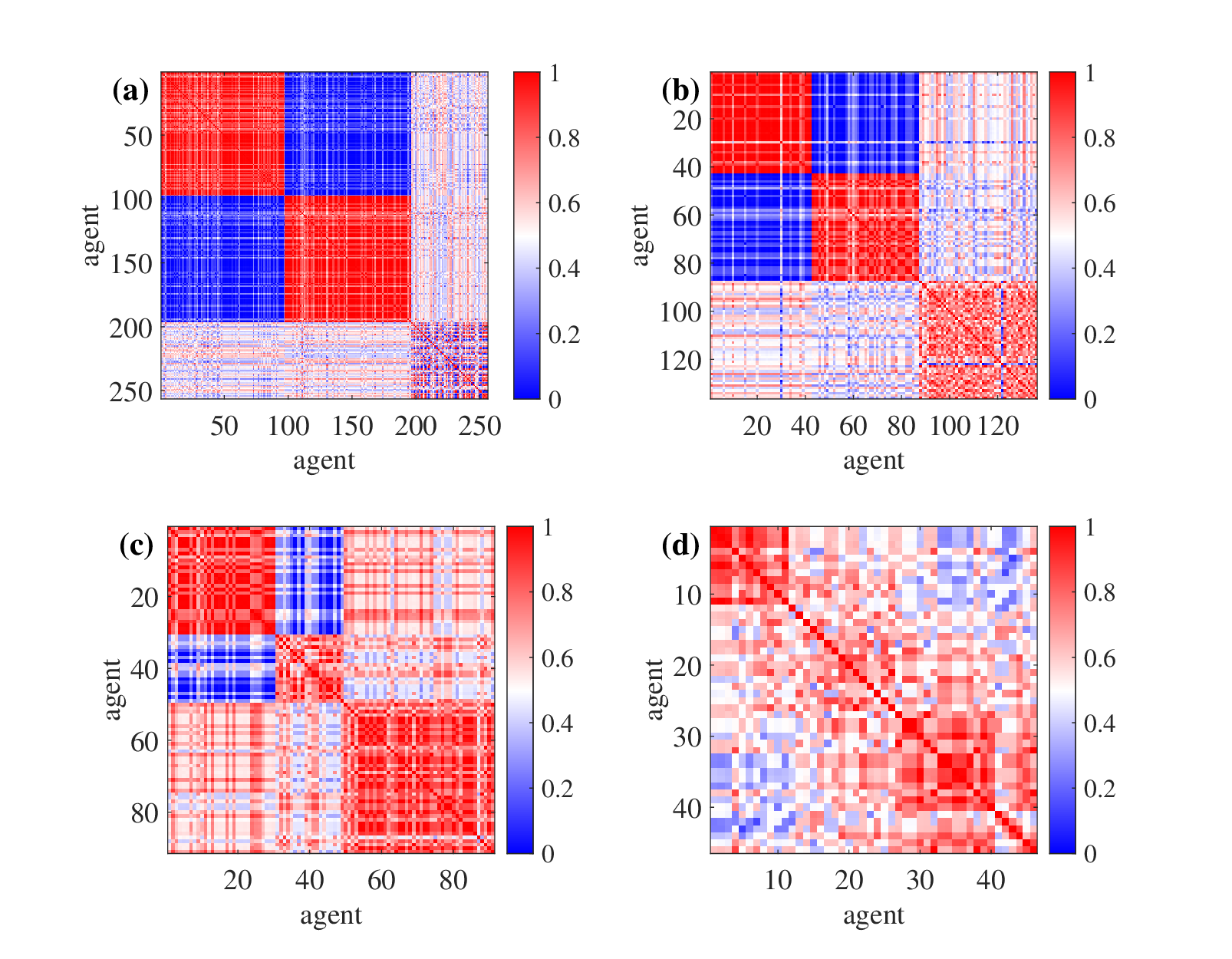}
	\caption{ (Color online) {\bf K-means clustering analysis for Q-subpopulation under the synchronization.} Panels (a-d) show the K-means clustering analysis results based on the synchronization between any pair of agents $i$ and $j$ within Q-subpopulation. When $f_{c}$ is far below the transition point $f_{c}^{*}$, the results indicate that the Q-subpopulation always consists of two clusters, $\mathcal{C}^{\text{\RNum{1}}}_{q}$ and $\mathcal{C}^{\text{\RNum{2}}}_{q}$, which participate in the intra-synergy within the Q-subpopulation.
	In contrast, the C-subpopulation does not have such clusters [See Fig.~\ref{fig:C_clustering}]. The internal synergy will fade away with the increase of $f_{c}$. In the results, the setup of number of cluster is $K = 3$. In (a-d), the fractions of C-population are $f_{c} = 0.15$, $0.55$, $0.7$ and $0.85$. 
	The learning parameters $(\alpha, \gamma, \epsilon)$, memory length $m$, size of strategy base $|\mathcal{B}|$ and system size $|\mathcal{N}|$ are set as those in Fig.~\ref{fig:volatility} by default. 
	}\label{fig:clustering}
\end{figure}

To further investigate the forms of intra-synergy and inter-synergy from the perspective of the Q-subpopulation, we initially investigate the synchronization between any agents $i$ and $j$ within the Q-subpopulation $\mathcal{N}_q$. Based on the action time series, 
the synchronization between $i$ and $j$ is defined as 
\begin{align}\label{eq:synchronization_factor}
	\sigma_{q}^{i,j} &:= 1 - \bar{d}_{H_{q}}(\bm{a}^i,\bm{a}^{j}) \nonumber\\
	 &= 1 - \frac{\sum\limits_{\tau=t_0}^{T} |a^{i}(\tau)-a^{j}(\tau)|}{T-t_0}, 
\end{align}
where $\bar{d}_{H_{q}}(\bm{a}^i,\bm{a}^{j})$ denotes the average Hamming distance between the time series $\bm{a}^i$ and $\bm{a}^{j}$ for $i$ and $j$. 
Then, we perform K-means clustering analysis to the matrix $\bm{\sigma}_{q}$ of $\mathcal{N}_q$. Intuitively, the number of cluster is set as $K = 3$. The clusters derived from the Q-population
are designated as $\mathcal{C}_{q}^{\text{\RNum{1}}}$, $\mathcal{C}_{q}^{\text{\RNum{2}}}$, $\mathcal{C}_{q}^{\text{\RNum{3}}}$. The results of K-means clustering analysis~\cite{mcqueen1967some} for Q-subpopulation are shown in Fig.~\ref{fig:clustering}.

As shown in Fig.~\ref{fig:clustering}(a - b), when $f_{c}$ is far below the transition point $f_{c}^*$, there is consistently two intra-synchronization (intra-sync) clusters of the same magnitude in the Q-subpopulation. Still, these synchronized clusters present inter-anti-synchronization (inter-anti-sync) between them. Without loss of generality, we appoint these clusters as $\mathcal{C}_{q}^{\text{\RNum{1}}}$ and $\mathcal{C}_{q}^{\text{\RNum{2}}}$, respectively. 
The intra-synchronization and the inter-anti-synchronization suggest that agents within the same cluster, be it $\mathcal{C}_{q}^{\text{\RNum{1}}}$ or $\mathcal{C}_{q}^{\text{\RNum{2}}}$, are inclined to access the same resource. In contrast, agents from $\mathcal{C}_{q}^{\text{\RNum{1}}}$ and $\mathcal{C}_{q}^{\text{\RNum{2}}}$ respectively, consistently choose opposite resources to access. The results suggest that the intra-synergy within the Q-subpopulation stems from the intra-synchronization of $\mathcal{C}_{q}^{\text{\RNum{1}}}$ and $\mathcal{C}_{q}^{\text{\RNum{2}}}$ as well as the inter-anti-synchronization between them. Therefore, we abbreviate $\mathcal{C}_{q}^{\text{\RNum{1}}}$ and $\mathcal{C}_{q}^{\text{\RNum{2}}}$ as \emph{intra-synergy clusters} (\emph{IS-clusters}).
In the meantime, the inter-subpopulation synergy is manifested in the synergy between $\mathcal{C}_{q}^{\text{\RNum{3}}}$ and the C-subpopulation. Consequently, we abbreviate $\mathcal{C}_{q}^{\text{\RNum{3}}}$ as the \emph{inter-synergy cluster} (\emph{ES-cluster}).
Additionally, another fascinating phenomenon is that as $f_{c}$ increases, the IS-clusters gradually shrink. In contrast, the ES-cluster expands and undergoes a transition from a disordered state to a synchronous one. This means the fraction $f_{c}$
determines the proportions of the IS-clusters and the ES-cluster within the Q-subpopulation. 

As $f_{c}$ further increases towards $f_{c}^*$, the IS-clusters $\mathcal{C}_{q}^{\text{\RNum{1}}}$ and $\mathcal{C}_{q}^{\text{\RNum{2}}}$ continue to shrink and their sizes turn asymmetrical [see Fig.~\ref{fig:clustering}(c)]. In addition, the anti-synchronization between them weakens, and instead, both of IS-clusters gradually start to synchronize with the ES-cluster $\mathcal{C}_{q}^{\text{\RNum{3}}}$. Furthermore, the intra-sync within $\mathcal{C}_{q}^{\text{\RNum{3}}}$ is further enhanced. These changes suggest that within the Q-subpopulation, intra-synergy gradually gives way to inter-synergy, ultimately resulting in complete inter-synergy in the form of synchronization at the transition point $f_{c}^*$. However, with the further increase of $f_{c}$ and exceeds $f_{c}^*$, the Q-subpopulation fails to be partitioned into distinct clusters via K-means clustering analysis [see Fig.~\ref{fig:clustering} (d)]. This indicates that 
both the intra-synergy and inter-synergy will be disrupted as long as the fraction of the Q-subpopulation is excessively low.

\subsubsection{Synchronization and anti-synchronization}\label{subsubsec:syn_antisyn}
\begin{figure*}[htbp!] 
	\centering 
	\includegraphics[width=0.90\textwidth]{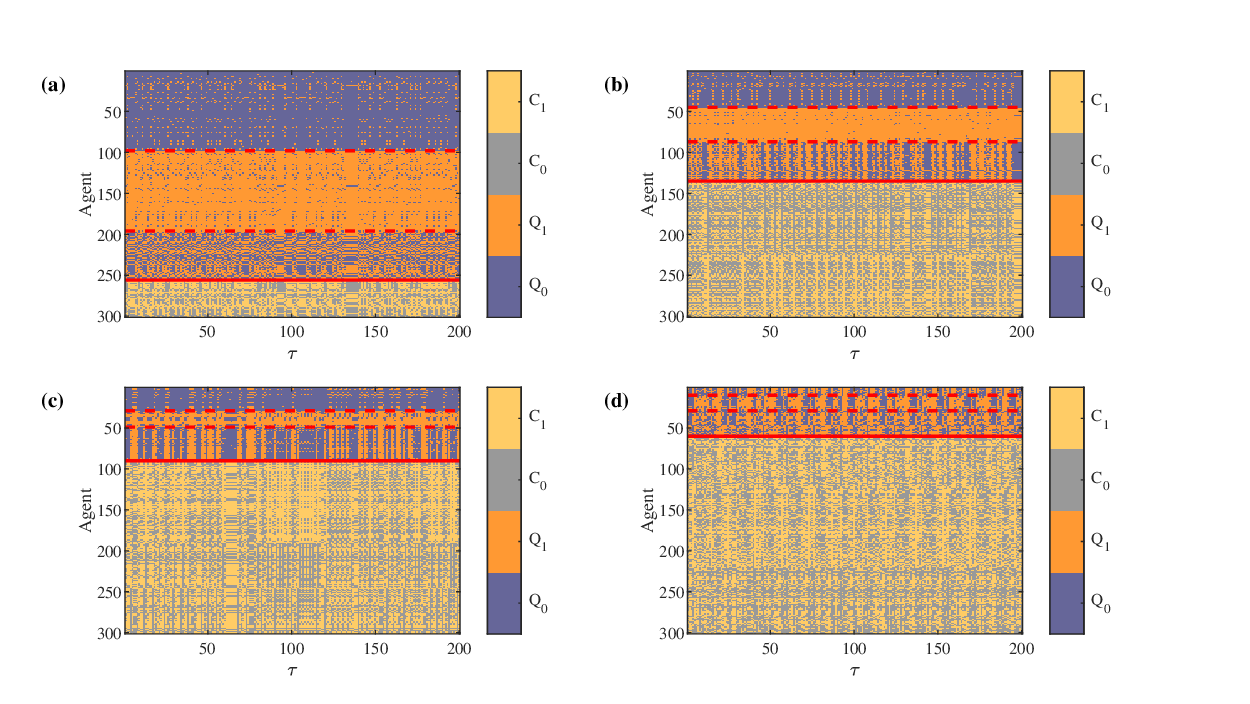}
	\caption{{\bf Time series of actions of agents within different subpopulations.} In panels (a) to (d), the boundary separating the Q-subpopulation from the C-subpopulation is indicated by a solid red line. Meanwhile, the boundaries between the three clusters $\mathcal{C}^{\text{\RNum{1}}}_{q}$, $\mathcal{C}^{\text{\RNum{2}}}_{q}$, and $\mathcal{C}^{\text{\RNum{3}}}_{q}$ within the Q - subpopulation are marked with two dashed lines. In (a - b), quasi-static intra-synchronization is seen in IS-clusters $\mathcal{C}^{\text{\RNum{1}}}_{q}$ and $\mathcal{C}^{\text{\RNum{2}}}_{q}$. In (c - d), as $f_{c}$ increases, the quasi-static intra-synchronization is gradually replaced by dynamic one and finally disappears. Also, (a - c) show that agents in $\mathcal{C}^{\text{\RNum{3}}}_{q}$, the ES-cluster, shift from a disordered to a dynamically-synchronized state.  
	In (a-d), the fractions of C-population are $f_{c} = 0.15$, $0.55$, $0.7$ and $0.85$, respectively, which correspond to (a-d) in Fig.~\ref{fig:clustering}. The learning parameters $(\alpha, \gamma, \epsilon)$, memory length $m$, size of strategy base $|\mathcal{B}|$ and system size $|\mathcal{N}|$ are set as those in Fig.~\ref{fig:volatility} by default. 
	} 
	\label{fig:timeseries_action}
\end{figure*}
To explore the dynamics of synchronization, we present the time series of agents of clusters under K-means clustering analysis in Fig.~\ref{fig:timeseries_action}. The results in (a) and (b) indicate that when $f_{c}$
is much lower than the transition point $f_{c}^*$, the intra-synchronization of IS-clusters is quasi-static, i.e., agents within $\mathcal{C}^{\text{\RNum{1}}}_q$ or $\mathcal{C}^{\text{\RNum{2}}}_q$ maintain their resource selections in a quasi-static manner. In addition, with the increase of $f_{c}$, the ES-cluster $\mathcal{C}^{\text{\RNum{3}}}_q$ gradually expands. Different from the quasi-static synchronization of  $\mathcal{C}^{\text{\RNum{1}}}_q$ and $\mathcal{C}^{\text{\RNum{2}}}_q$, $\mathcal{C}^{\text{\RNum{3}}}_q$ transitions from a disordered state to a dynamically-synchronized one. In other words, agents within $\mathcal{C}^{\text{\RNum{3}}}_q$ tend to choose the same resource, yet their choices evolve over time. 

As $f_{c}$ further increases and approaches $f_{c}^*$, the initially equally-sized IS-clusters $\mathcal{C}^{\text{\RNum{1}}}_q$ and $\mathcal{C}^{\text{\RNum{2}}}_q$ become unequal [see Fig.~\ref{fig:timeseries_action}(c)]. Meanwhile, the intra-synchronization of these two clusters gradually changes from static to dynamic, particularly for the smaller one. Additionally, the inter-synchronization between $\mathcal{C}^{\text{\RNum{1}}}_q$ and $\mathcal{C}^{\text{\RNum{3}}}_q$, as well as that between $\mathcal{C}^{\text{\RNum{2}}}_q$ and $\mathcal{C}^{\text{\RNum{3}}}_q$, increases slightly. Additionally, the results in (a - c) demonstrate the action preferences of agents within $\mathcal{C}^{\text{\RNum{3}}}_q$ and those of agents within $\mathcal{N}_{c}$ are becoming increasingly opposite as $f_{c}$  increases. This further supports that $\mathcal{C}^{\text{\RNum{3}}}_q$ plays a main role in the inter-subpopulation synergy between $\mathcal{N}_{c}$ and $\mathcal{N}_{q}$.
When $f_{c}$ exceeds $f_{c}^*$, the boundaries between the clusters blur, and the entire Q-subpopulation maintains a low level of synchronization. Furthermore, a characteristic time emerges during the synchronization evolution process.

In Appendix \ref{sec:app_syn_antisyn}, the analysis reveals that for the Q-subpopulation to achieve the optimal intra-synergy through IS-clusters, two conditions must be met:
\begin{enumerate}
\item The volatility of $\mathcal{C}^{\text{\RNum{1}}}_q$ and $\mathcal{C}^{\text{\RNum{2}}}_q$, denoted as $\psi_{qq}$, approaches $0$.
\item The expected number of agents entering resource 1 in $\mathcal{C}^{\text{\RNum{1}}}_{q}$ and $\mathcal{C}^{\text{\RNum{2}}}_{q}$ is equal to half of the total number of people in the two clusters, i.e.,
\begin{align}
	\mathbb{E}(N^{\text{\RNum{1}}}_{q_1}(\tau)+N^{\text{\RNum{2}}}_{q_1}(\tau))=(|\mathcal{C}^{\text{\RNum{1}}}_{q}| + |\mathcal{C}^{\text{\RNum{2}}}_{q}|)/2.  \nonumber
\end{align}
\end{enumerate}
There are two approaches to fulfill these two conditions. Firstly, both IS-clusters can be quasi-statically intra-synchronized, have the same size, and approximately exhibit inter-anti-synchronization with each other [see Eq.~\eqref{app_eq:conditions} under $\langle\sigma^{\text{\RNum{1},\RNum{2}}}_q\rangle\approx 0$]. Secondly, both IS-clusters can be dynamically intra-synchronized, but they differ in size and display weak inter-anti-synchronization. Figure.~\ref{fig:timeseries_action} illustrates that the former situation occurs when $f_{c}$ is low [see (a-b)]. Conversely, the latter phenomenon takes place when $f_{c}$ is high but still below the transition point [see (c)]. 

In summary, the IS-clusters $\mathcal{C}_{q}^{\text{\RNum{1}}}$ and $\mathcal{C}_{q}^{\text{\RNum{2}}}$ play a crucial role in suppressing the intra-volatility of the Q-subpopulation partly through intra-synchronization and inter-synchronization. In contrast, the ES-cluster $\mathcal{C}_{q}^{\text{\RNum{3}}}$ contributes to suppressing the volatility of the C-subpopulation by participating in the inter-subpopulation synergy. Thus, for low $f_{c}$, the intra-synergy within the Q-subpopulation is the dominant factor in resource allocation optimization. While, as $f_{c}$ increases, this role is taken over by the inter-subpopulation synergy between $\mathcal{N}_{q}$ and $\mathcal{N}_{c}$. Additionally, $\mathcal{C}^{\text{\RNum{3}}}_q$ also grows in size and eventually comes to dominate the Q-subpopulation. However, once $f_{c}$ exceeds the transition point $f_{c}^{*}$, even when $\mathcal{C}^{\text{\RNum{3}}}_q$ takes up the entire Q-subpopulation, the inter-synergy is unable to suppress the volatility of the C-subpopulation. This results in the emergence of a disordered phase.
\subsubsection{Freeze effect and momentum strategy}
\begin{figure*}[htbp!] 
	\centering 
	\includegraphics[width=0.95\textwidth]{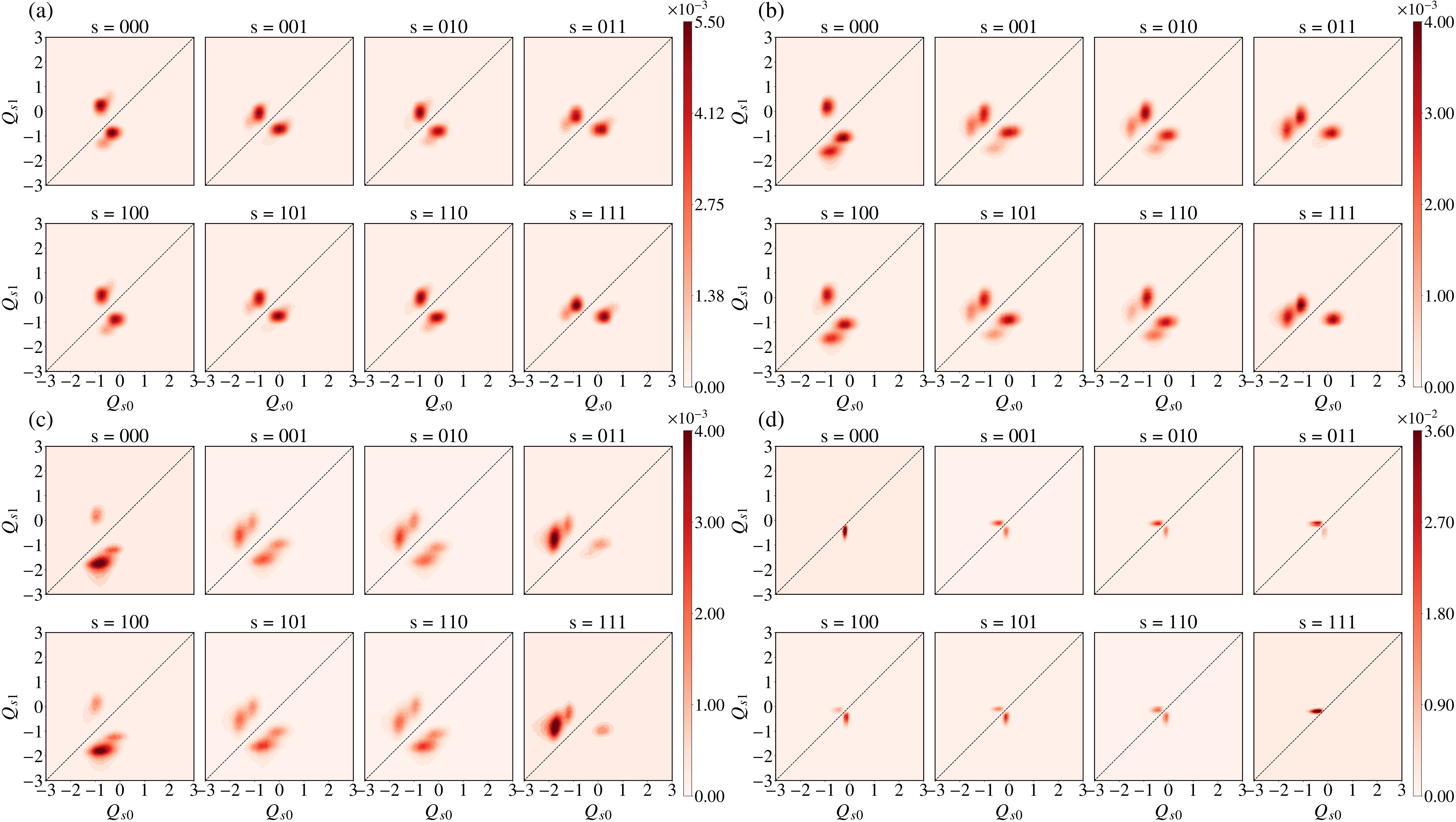}
	\caption{{\bf Distributions of state-action values for competing actions across different states.} (a-d) show the distributions of Q-values for competing actions across various states within an ensemble composed of $20$ runs. The momentum strategy is observed in (b-d). In (a-d), the fractions of C-population are $f_{c} = 0.15$, $0.55$, $0.7$ and $0.85$, which correspond to (a-d) of Fig.~\ref{fig:clustering}. The learning parameters $(\alpha, \gamma, \epsilon)$, memory length $m$, size of strategy base $|\mathcal{B}|$ and system size $|\mathcal{N}|$ are set as those in Fig.~\ref{fig:volatility} by default. } 
	\label{fig:Qtables}
\end{figure*}

The freeze effect, as previously demonstrated in relevant research~\cite{challet2004minority,coolen2005mathematical,coolen2001dynamical,moro2004minority}, plays a pivotal role in optimizing resource allocation in the MG. Consequently, we also direct our attention to this effect within the Q-subpopulation. For agents in $\mathcal{N}_{q}$, the concept of ``freeze'' means that the cognitively optimal action across different states stays constant. The robustness of this freeze against noise is determined by the gap in Q-values between competing actions of different states. As a result, we present the probability density function (PDF) of the Q-values for competing actions in different states in Fig.~\ref{fig:Qtables}. Within this space, should the Q-values be distributed on the diagonal line (where $Q_{s,0} = Q_{s,1}$), the agents have no preference between entering resource $1$ and resource $0$. In contrast, when the Q-values are distributed in the upper-left of the diagonal (where $Q_{s,1} > Q_{s,0}$), the agents show a preference for entering resource $1$ at state $s$; when they are in the lower-right (where $Q_{s,1} < Q_{s,0}$), the preference shifts towards resource $0$ at the same state. Evidently, the robustness of this frozen preference against noise is positively correlated with the distance between agents' competing Q-values and the diagonal.

When the fraction $f_{c}$ is low, within PDF corresponding to any state, two distinct aggregations can be observed. One is distributed in the upper-left of the diagonal, and the other in the lower-right. Significantly, these aggregations are quite far away from the diagonal [see Fig.~\ref{fig:Qtables} (a)].
It is obvious that these well-defined aggregations correspond to the distributions of competing Q-values of agents within the IS-clusters $\mathcal{C}_{q}^{\text{\RNum{1}}}$ and $\mathcal{C}_{q}^{\text{\RNum{2}}}$. And the considerable distance separating these aggregations from the diagonal indicates that the preference of these agents has a strong robustness against noise. This, in turn, ensures the stability of the intra-synergy within the Q-subpopulation. Moreover, it is notable that there are also some blurry aggregations distributed in space. This is especially evident in states where a certain resource has had continuous winning in the recent rounds, like $s_{0}(000)$, $s_{7}(111)$, $s_{3}(011)$ and $s_{4}(100)$. A justifiable conjecture is that these unclear clusters are Q-value distributions of agents within ES-cluster $\mathcal{C}_{q}^{\text{\RNum{3}}}$ of Q-subpopulation. 

As $f_{c}$ keeps rising yet stays below the transition point $f_{c}^{*}$, the well-defined clusters observed at low $f_{c}$ progressively turn blurry. In contrast, the initially blurry aggregations become clearly distinguishable [see Fig.~\ref{fig:Qtables}(b-c)].
Drawing on the analysis in Sec.~\ref{subsubsec:Intra_synergy}, the results further corroborate the previous conjecture that the aggregations, being indistinct at low $f_{c}$ but becoming distinct, correspond to the distributions of the competing Q-values of the agents within $\mathcal{C}_{q}^{\text{\RNum{3}}}$. Moreover, in the states where a particular resource consistently wins in recent rounds, such as $s_{0}$, $s_{7}$, $s_3$ and $s_4$, the aggregations of $\mathcal{C}_{q}^{\text{\RNum{3}}}$ exist only on one side of the diagonal. This result demonstrates agents within the ES-cluster $\mathcal{C}_{q}^{\text{\RNum{3}}}$ exhibit a clear preference for the resource that has achieved consecutive successes recently. This phenomenon implies the emergence of the well-known ``\emph{momentum strategy}'',  chasing rising prices and selling falling prices~\cite{chan1996momentum,menkhoff2012currency}, emerge in $\mathcal{C}_{q}^{\text{\RNum{3}}}$. However, the competing Q-values for the aggregation of $\mathcal{C}_{q}^{\text{\RNum{3}}}$ consistently stays lower than those for the other aggregation located on the same side of the diagonal as the aggregation of $\mathcal{C}_{q}^{\text{\RNum{3}}}$. The result suggests that the long-term reward for agents within the ES-cluster is lower than that for agents within the IS-clusters. This finding further indicates that the momentum strategy benefits from optimizing resource allocation within the population. However, agents adopting these strategies receive a lower long-term reward compared to the others, rather than a higher one.
At last, all aggregations remain far from the diagonal. This indicates that the robustness of the freeze effect against noise persists regardless of whether the agents belong to the IS-clusters $\mathcal{C}_{q}^{\text{\RNum{1}}}$ and $\mathcal{C}_{q}^{\text{\RNum{2}}}$, or the ES-cluster $\mathcal{C}_{q}^{\text{\RNum{3}}}$.

As $f_{c}$ further increases and is above $f_{c}^{*}$, the aggregations of Q-value distribution of all agents on the same side of the diagonal merge together and approach the diagonal [see Fig.~\ref{fig:Qtables} (d)]. The result indicates that all agents within $\mathcal{N}_{q}$ possess similar cognition, and the IS-clusters and the ES-cluster merge with each other. Additionally, the robustness of the freeze effect against noise nearly vanishes. However, the results of $s_{0}$, $s_{7}$, $s_{3}$ and $s_{4}$ demonstrate that all agents in the Q-subpopulation are still attempting to prevent the long-term under-utilization of a resource via momentum strategy. This implies that the entire Q-subpopulation engages in inter-synergy, while intra-synergy based on synchronization and anti-synchronization diminishes.

\subsubsection{Effect of the momentum strategy}
\begin{figure}[htbp!] 
	\centering 
	\includegraphics[width=0.485\textwidth]{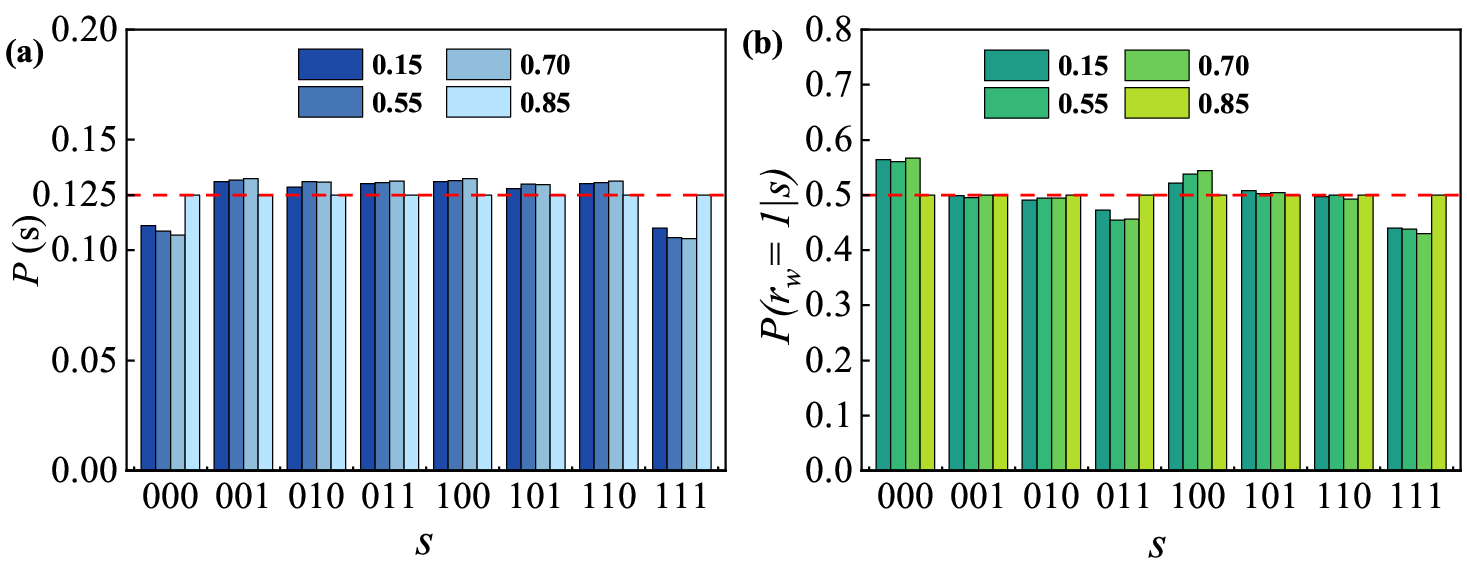}
	\caption{{\bf Distribution of states and the conditional winning probability distribution of a specific resource for different given states.} (a) shows the probability distribution $p(s)$ corresponding to various levels of $f_{c}$. While (b) exhibits the conditional winning probability $p(r_{w} = 1|s)$ given different states corresponding to various levels of $f_{c}$. In (a) and (b), the red dotted lines respectively mark $p(s)$ and $p(r_{w} = 1|s)$ of the random-choice system, which serve as benchmarks. Each result is derived from an ensemble consisting of $20$ runs. The learning parameters $(\alpha, \gamma, \epsilon)$, memory length $m$, size of strategy base $|\mathcal{B}|$ and system size $|\mathcal{N}|$ are set as those in Fig.~\ref{fig:volatility} by default.}
	\label{fig:ps_ps1}
\end{figure}

To delve into the momentum strategy of ES-cluster $\mathcal{C}^{\text{\RNum{3}}}_q$ influences resource allocation, we investigate the probability of a state $s_{\mu}$ and the conditional winning probability of a specific resource $r$ given state $s_\mu$ as Fig.~\ref{fig:ps_ps1} (a-b) shows. These probabilities are defined as 
\begin{align}\label{eq:p_state}
	p(s_\mu) :=  \frac{\sum\limits_{\tau = t_0}^{T}\mathbbm{1}_{s(\tau) = s_\mu}}{T-t_0}, 
\end{align} 
and 
\begin{align}\label{eq:p_condition}
	p(r_w = r|s_\mu) :=  \frac{\sum\limits_{\tau = t_0}^{T}\mathbbm{1}_{{s(\tau) = s_\mu},{r_w(\tau) = r}}}{\sum\limits_{\tau = t_0}^{T}\mathbbm{1}_{s(\tau) = s_\mu}}. 
\end{align}  

The results in (a) show that, when $f_{c}$ is lower than the transition point $f_{c}^{*}$, $p(s)$
for $s_{0}(000)$ or $s_{7}(111)$ are decreased as $f_{c}$ increases.  Moreover, these values are consistently and significantly lower than $1/|\mathcal{S}|$, the probability of these states in the random choice model. 
This implies that the momentum strategy of agents within the ES-clusters $\mathcal{C}_{q}^{\text{\RNum{3}}}$ effectively averts the long-term under-utilization of a particular resource. Consequently, it further refines the resource allocation within the population. This phenomenon also explains why the level of resource allocation improves as the fraction of $\mathcal{C}_{q}^{\text{\RNum{3}}}$ in the Q-subpopulation increases. However, as $f_{c}$ further increases and exceeds $f_{c}^{*}$, $p(s)$ for $s_{0}(000)$ or $s_{7}(111)$ suddenly approach $1/|\mathcal{S}|$. This finding suggests that when only a small number of agents adopt momentum strategies, they are unable to prevent the long-term under-utilization of a particular resource, and further supports the phase transition at $f_{c}^{*}$ is of the first order. 

In (b), we can observe $p(r_w = 0|s_\mu)$ given state $s_0(000)$ or $s_{4}(100)$ is lower than $1/2$, which is the corresponding $p(r_w = 0|s_\mu)$ in random choice game. Likewise, $p(r_w = 1|s_\mu)$ given $s_7(111)$ or $s_{3}(011)$ is also lower than $1/2$. This finding suggests that while the momentum strategies of agents can prevent the long-term under-utilization of any resource within the system, it will, however, lead to trend reversals and a decrease in their winning probability. The result further elucidates the reason behind the phenomenon that the long-term reward for agents in the ES-cluster is lower than that for agents in the IS-clusters in Fig.~\ref{fig:Qtables}.

\subsection{Analysis of C-subpopulation}\label{sec:analysis_csubpopulation}
\begin{figure*}[htbp!] 
	\centering 
	\includegraphics[width=0.47\textwidth]{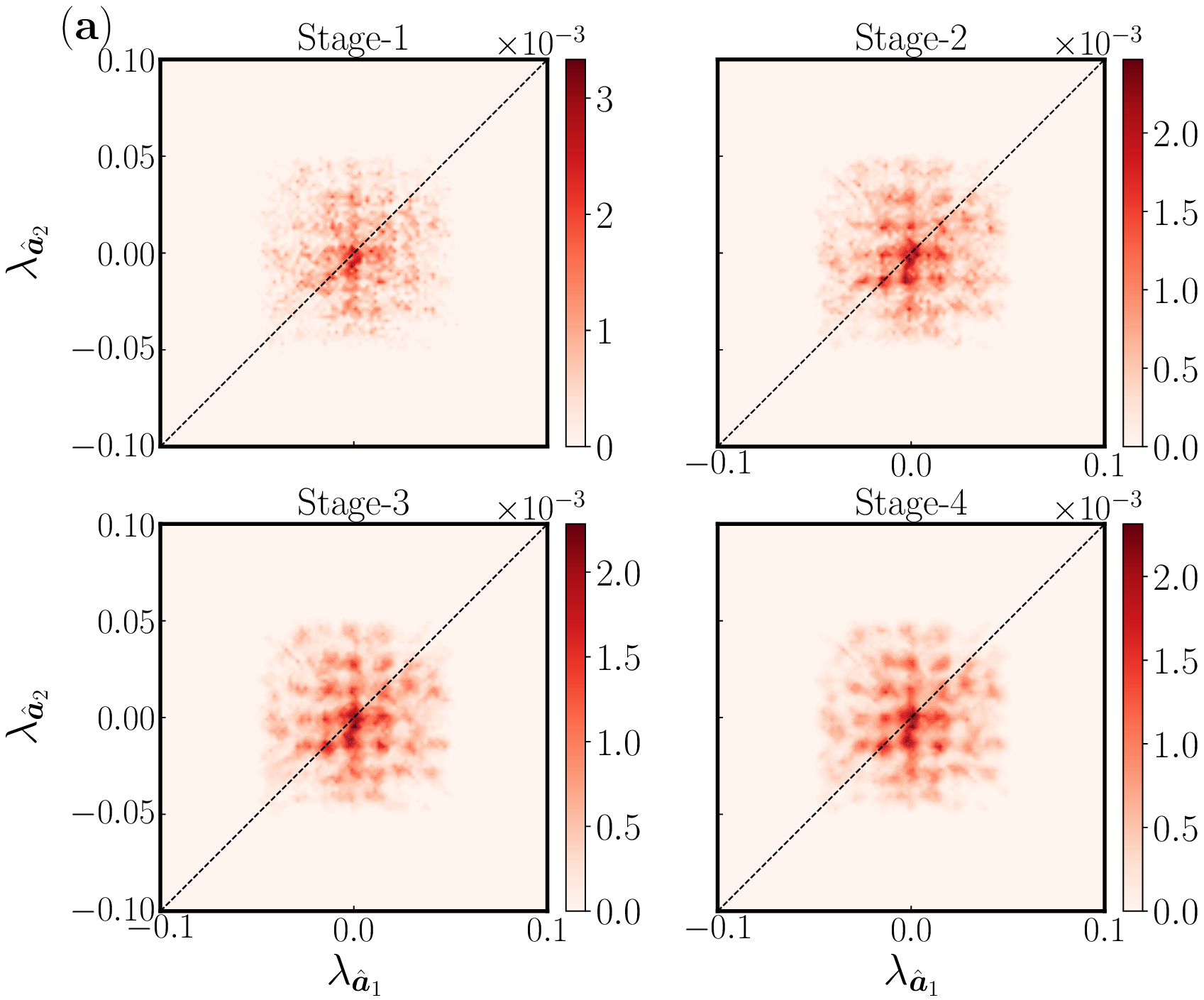}
	\hfill
	\includegraphics[width=0.47\textwidth]{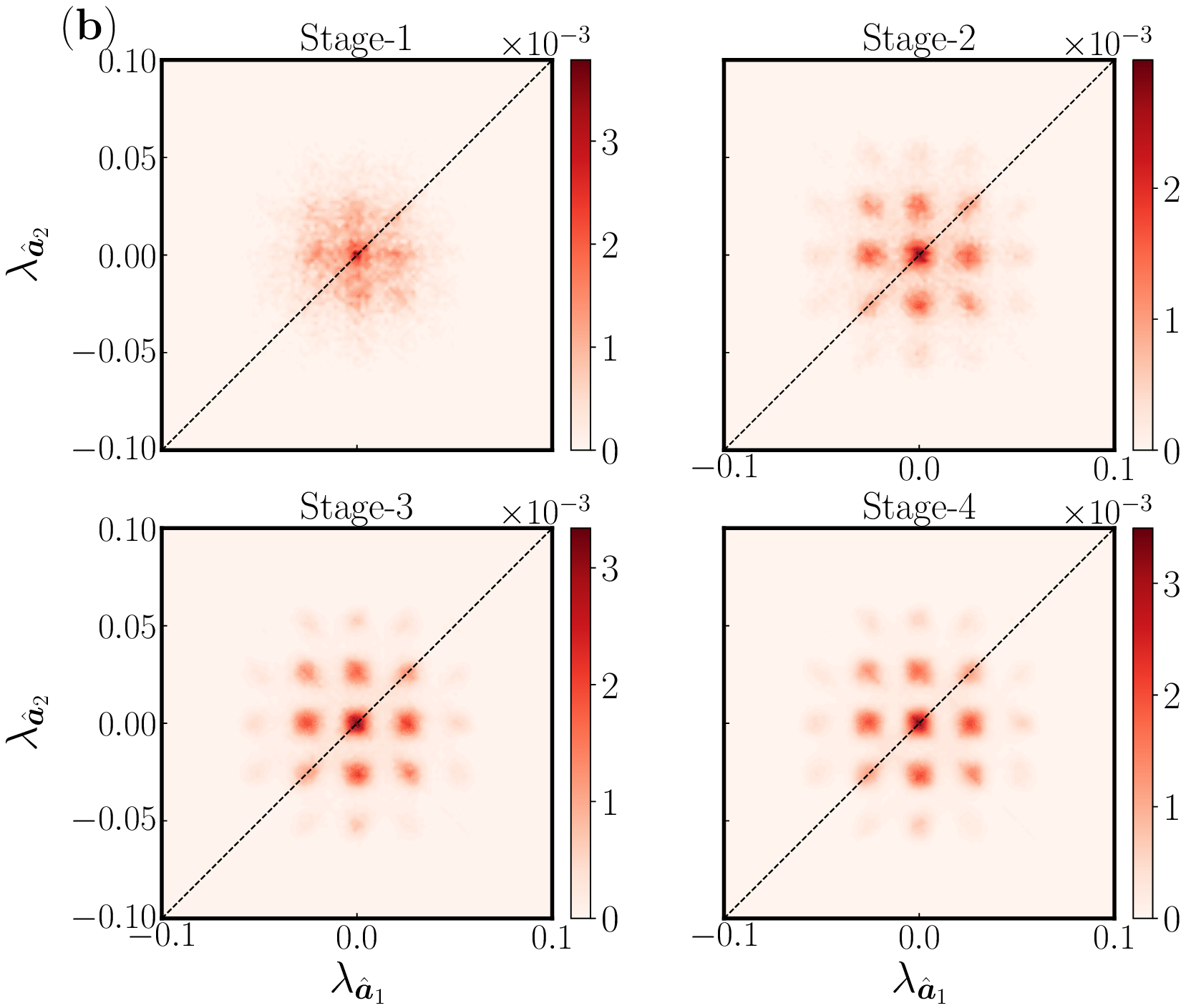}
	\\
	\includegraphics[width=0.47\textwidth]{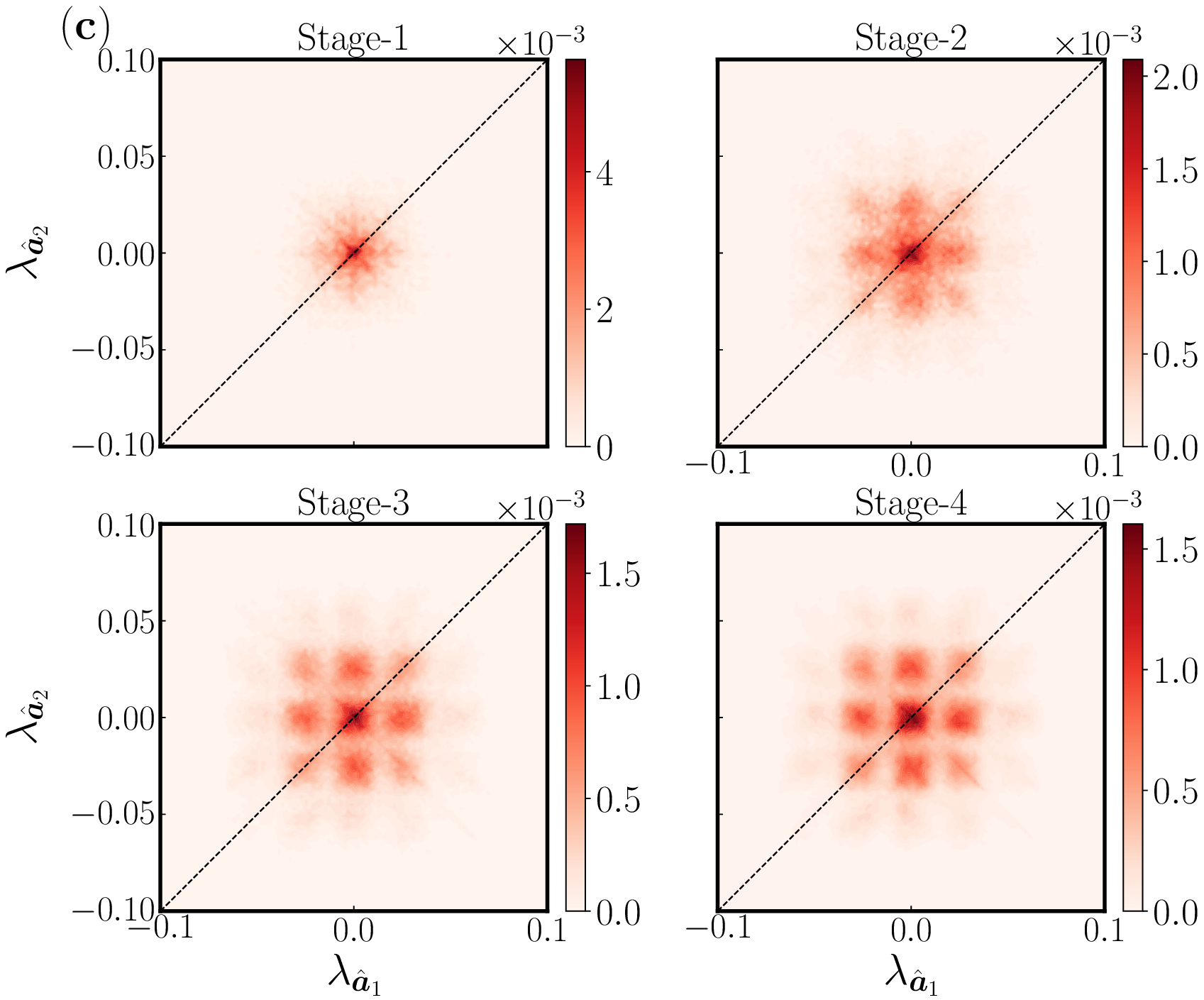}
	\hfill
	\includegraphics[width=0.47\textwidth]{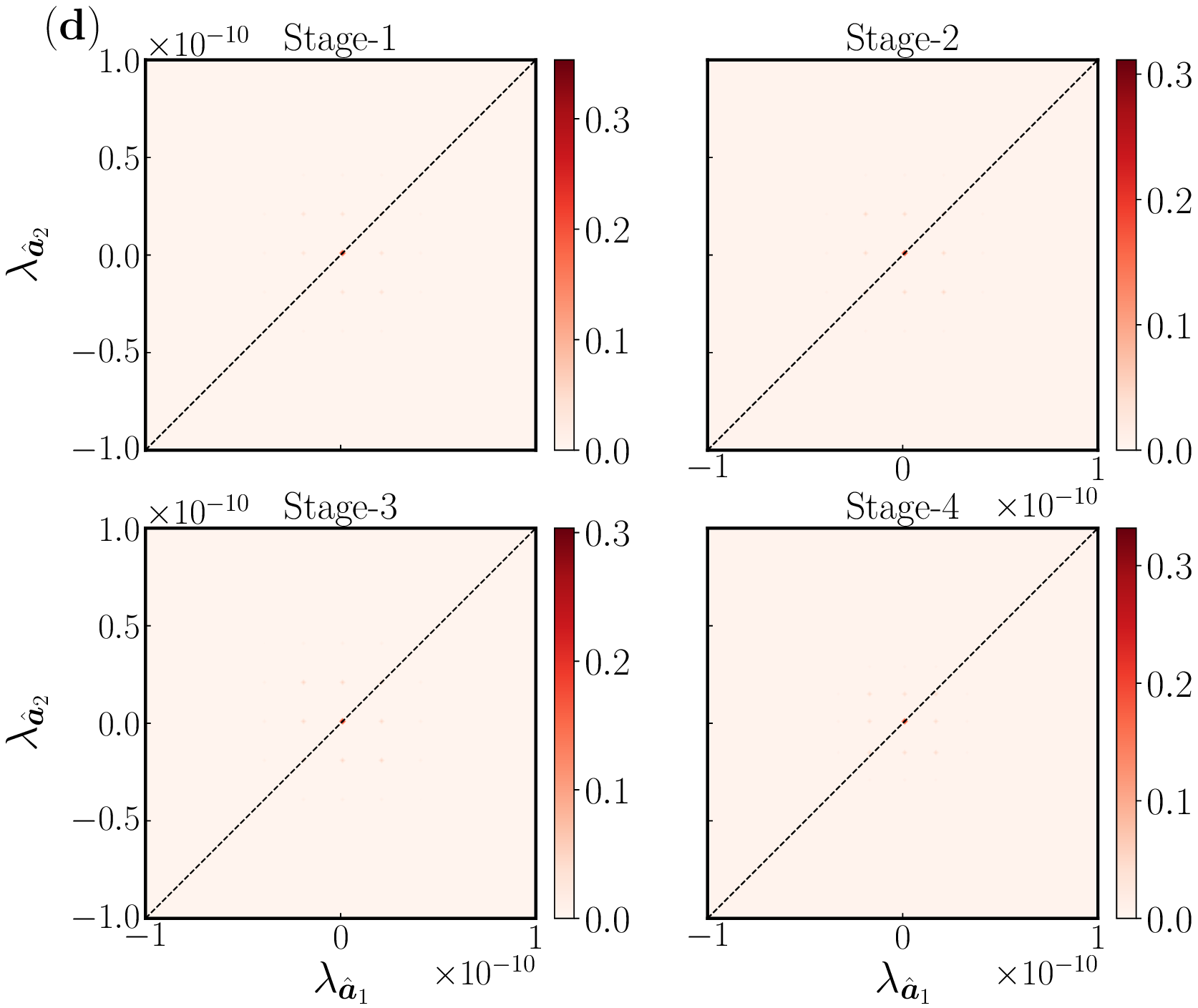}
	\caption{{\bf The distribution of the growth rates of competing strategies’ scores in C-subpopulation.} (a-d) show the distributions of growth rates for competing strategies’ scores across different stages. The $2\times 10^{7}$ steps are divided into 
		$200$ unit intervals and each one includes $\Delta\tau = 10^5$ steps. Stage-1 to Stage-4 in each panel contain  $\tuple{7, 21, 57, 115}$ unit intervals in sequence. In the figure (a-d), the fractions of C-population are $f_{c} = 0.15$, $0.55$, $0.7$ and $0.85$ respectively, which are correspond to (a-d) of Fig.~\ref{fig:clustering}. the result of each panel is derived from an ensemble consisting of $20$ runs. The learning parameters $(\alpha, \gamma, \epsilon)$, memory length $m$, size of strategy base $|\mathcal{B}|$ and system size $|\mathcal{N}|$ are set as those in Fig.~\ref{fig:volatility} by default.} 
	\label{fig:growth_rate}
\end{figure*}
For an agent in the C-subpopulation, the term ``freeze'' implies that the cognitively superior strategy selected from the agent’s strategy base $\mathcal{B}$ remains unaltered~\cite{challet2004minority,moro2004minority}. The robustness of this freeze against noise depends on the score gap between competing strategies.
However, in contrast to the Q-values of the Q-subpopulation, the scores of the C-subpopulation diverge rather than converge. As a result, we focus on the PDF of the growth rates of competing strategies’ scores during different stages, rather than the PDF of the scores themselves. The growth rate of any strategy over a unit interval $\Delta \tau$ at $\tau$ is defined as
\begin{align}\label{eq:growth_rate}
	\lambda_{\hat{\bm{a}}}:= \frac{\text{score}(\hat{\bm{a}}, \tau) -  \text{score}(\hat{\bm{a}}, \tau - \Delta \tau)}{\Delta \tau}. 
\end{align}
The PDF of the growth rates of competing strategies’ scores during different stages are shown in Fig.~\ref{fig:growth_rate}. Evidently, the freezing rate is positively correlated with the distance between the distribution of the growth rates of agents’ competing strategies and the diagonal.
 
During the Stage-$1$, when $f_{c}$ is lower than the transition point $f_{c}^{*}$, the PDF of the growth rates show that the scores change relatively slowly and irregularly over time [see Fig.~\ref{fig:growth_rate}(a-c)]. This phenomenon results from the agents’ disordered and unfrozen actions caused by their chaotic exploration. Additionally, as $f_c$ increases, the rate of change of the scores decreases. This indicates that the freezing rate is negatively correlated with $f_c$. However, once the system enters Stage-$2$, some blurry aggregations emerge, especially at medium $f_c$ as (c) shows. This marks the arrival of the initial frozen-tendency stage, causing some strategies with similar winning or losing probabilities over extended periods to tend to aggregate together in the PDF. As the system evolves further into Stage-$3$, the aggregations gradually become clearer because the approach to the substantially frozen stage causes changes in each strategy's score to stabilize. After the system is stable at Stage-4, the aggregations are similar to those in the third stage and hardly change any longer. However, different from medium and high $f_{c}$, the aggregations exhibit disorder for low $f_{c}$ [see (a)]. In addition, as $f_c$ approached $f_c^*$, the size of each aggregation swells, and the boundaries between them gradually melt away. This suggests that the frozen degree gradually weakens and the disorder increases near the transition point [see (d)]. 

Regardless of the stage, when $f_{c}$ is higher than the transition point $f_{c}^{*}$, the PDFs of the growth rates always center around the origin,  This implies that the long-term winning probability of each strategy approaches zero. As a result, neither of the two competing strategies can achieve consistent wins to the extent that it becomes a frozen strategy. Consequently, the system fails to reach the frozen stage and keeps disordered. The results in (a - d) indicate that the freezing effect plays a significant role in our DRLP-MG model, similar to the findings in previous works~\cite{challet2004minority,coolen2005mathematical,coolen2001dynamical,moro2004minority}.

\begin{figure}[htbp!] 
	\centering 
	\includegraphics[width=0.485\textwidth]{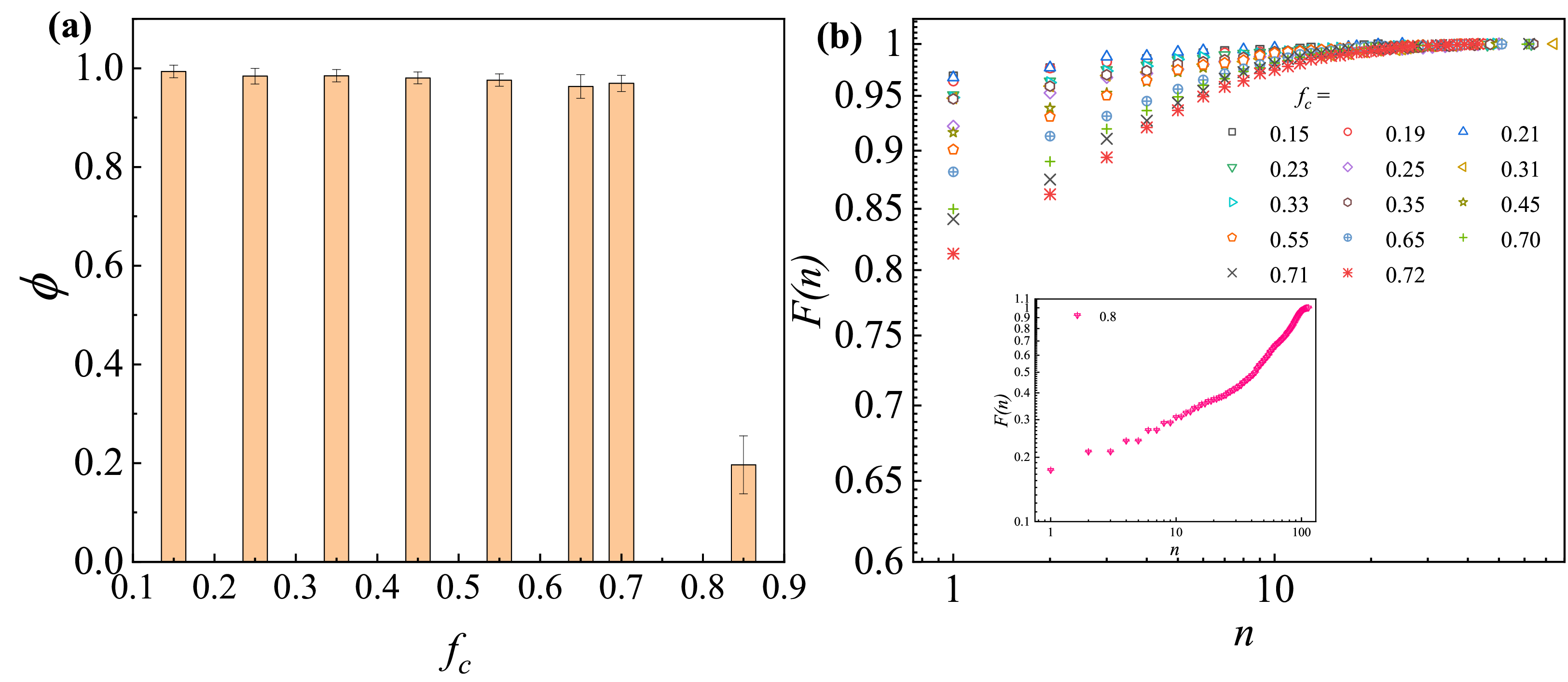}
	\caption{{\bf Two metrics in the C-population: the frozen ratio and the cumulative distribution of the number of optimal strategy switches for agents.} (a) shows the frozen ratio, denoted as $\phi$, corresponding to different values of $f_{c}$. When $f_{c}$ is lower than the transition point $f_{c}^{*}$, the ratio is extremely high. In contrast, when $f_{c}$ is higher than $f_{c}^{*}$, the ratio drops to a low level. The main panel of (b) depicts the cumulative distribution of the number of optimal strategy switches for agents, denoted as $F(n)$ at different values of $f_{c}$ when $f_{c}<f_{c}^{*}$. The inset shows the corresponding distribution when $f_{c}>f_{c}^{*}$. The result of each panel is derived from an ensemble consisting of $20$ runs. The learning parameters $(\alpha, \gamma, \epsilon)$, memory length $m$ and system size $|\mathcal{N}|$ are adopted by default as in Fig.~\ref{fig:volatility}.       
}
	\label{fig:frozen_cagents}
\end{figure}
Following this indication, we further investigate the ratio of frozen agents within the C-population. The frozen ratio $\phi$ is defined as follows
\begin{align}\label{eq:frozen_ratio}
	\phi := \sum_{i\in\mathcal{N}_{c}}\frac{\prod\limits_{k = 1}^{K} \mathbbm{1}_{\hat{\bm{a}}^{i}(\tau+k\Delta \tau)=\hat{\bm{a}}^{i}(\tau)}}{|\mathcal{N}_{c}|},
\end{align}
where $\hat{\bm{a}}^{i}(\tau)$ is the strategy which has the highest accumulated score in base $\mathcal{B}^i$ at $\tau$ as Eq.~\eqref{eq:classical_action} shows. Figure~\ref{fig:frozen_cagents} (a) depicts $\phi$ corresponding to different values of $f_{c}$.
The results show that, when $f_{c}$ is lower than transition point $f_{c}^*$, $\phi$ slightly decreases with the increase of $f_{c}$, yet it still stays close to $1$. However, once $f_{c}$ crosses $f_{c}^*$, $\phi$ drops rapidly. These results not only further support the previous indication but also imply that a small number of unfrozen agents within C-subpopulation might play a role in optimizing resource allocation.  

In Figure~\ref{fig:frozen_cagents} (b), we further provide the cumulative distribution function (CDF) of the number of switches of the optimal strategy for agents in C-subpopulation. Specifically, we define the number of such switches for any agent $i$ as
\begin{align}
	n_{\text{switch}}^{i} :=  \sum_{k = 1}^{K} \mathbbm{1}_{\hat{\bm{a}}^{i}(\tau + k\Delta\tau) \ne \hat{\bm{a}}^{i}(\tau + (k-1)\Delta\tau)}
\end{align}
and denote the CDF as $F(n) = P( n_{\text{switch}}<n)$. For a given $n$, $F(n)$ nearly always decreases with the increase of fraction $f_{c}$ when $f_{c}$ lies below $f_{c}^{*}$. The results further confirm the conclusion shown in Fig.~\ref{fig:growth_rate} that the frozen rate decreases with $f_{c}$. However, an exception occurs around $f_{c} = 0.25$, where
$F(n)$ first increases and then decreases as $f_{c}$ increases. Figures~(\ref{fig:clustering} - \ref{fig:timeseries_action}) and (\ref{fig:growth_rate}) have revealed that the synergy between subpopulations involves a transition from disorder to order as 
$f_{c}$ increases. This exception around $f_{c} = 0.25$ may indicate that this transition takes place near this point. The inset of the panel (b) also demonstrates that the agents within C-subpopulation are barely frozen when $f_{c}$ exceeds $f_{c}^{*}$.

\section{Discussion and Conclusion}\label{sec:summary} 
In this work, we present a model of Dual Reinforcement Learning Policies within the framework of the Minority Game (DRLP-MG) to investigate the synergy between these policies for optimizing resource allocation. In the model, the two policies are the classical policy~\cite{challet2004minority,moro2004minority} and the Q-learning policy~\cite{sutton2018reinforcement}, which are adopted by the C-subpopulation and the Q-subpopulation within the overall population, respectively. Based on the model, we discover that there exists a first-order phase transition as the fraction of C-population, denoted as 
$f_{c}$, increases. The optimal synergy occurs at the transition point $f_{c}^{*}$, at which resource allocation in the population is optimized. 
According to the K-means clustering analysis on the synchronization of Q-subpopulation, we find that the Q-subpopulation consists of two clusters involved in internal synergy (IS-clusters) within the Q-subpopulation and one cluster involved in external synergy (ES-cluster) between subpopulations when $f_{c}$ is below $f_{c}^{*}$. With the increase of $f_{c}$, the IS-clusters shrink while the ES-cluster expands, and the former will fade away when $f_{c}$ exceeds $f_{c}^{*}$. 

Moreover, the analysis of the distribution of state-action values for the Q-subpopulation significantly reveals that, solely through reinforcement learning, the classical momentum strategy in financial markets~\cite{chan1996momentum,menkhoff2012currency} emerges within the ES-cluster. This strategy contributes to resource allocation by preventing any resource from being under-utilized over the long term. However, the reversal of the leading resource’s trend caused by this strategy results in the agents within the ES cluster achieving lower long-term returns compared to other agents, instead of getting higher returns as happens in reality\cite{efficiency1993returns,jegadeesh2001profitability}. This difference exists because, unlike the real world, our model does not incorporate a delayed effect. In addition, our results also indicate that the condition for the emergence of the momentum strategy in our model is the presence of heterogeneity of learning granularity within the population, which may map to the real-world factors that give rise to the well-known classical momentum strategy in financial markets.

Lastly, our results prominently show that, akin to previous studies~\cite{challet2004minority,moro2004minority,zheng2023optimal}, the frozen effect still plays a vital role in resource allocation. Nevertheless, a certain fraction of unfrozen agents contributes to enhancing the synergy between subpopulations. Additionally, we offer mathematical analysis of the results, including the relationship between  resource allocation levels within the population and fluctuation of subpopulations, as well as the forms of synchronization and anti-synchronization of the IS-cluster in the Q-subpopulation.

Our research has shown that reinforcement learning policies with different granularities can form synergistic effects in resource allocation through the momentum strategy, yet several open questions remain. Firstly, while our model predicts the momentum strategy will yield lower returns, it’s unclear if adding a delay effect can transform it to generate higher returns in practic~\cite{jegadeesh2001profitability}. Secondly, the heterogeneity of the agent’s Q-table makes it nearly impossible to use the analysis in previous work and theoretically determine the exact phase-transition point~\cite{ding2023emergence,li2025cooperation,ding2024emergence}. Lastly, the computational complexity of DRLP-MG impedes the identification of the transition point via numerous simulations. Addressing these questions can guide future research and deepen our understanding of the synergistic effect of self-organized resource allocation from a reinforcement-learning perspective.

\section*{ACKNOWLEDGEMENTS}
We are supported by the Natural Science Foundation of China under Grants No. 12165014 and 12075144, and the Natural Science Basic Research Program of Shaanxi (Grant No.\ 2025JC-YBMS-019)
\section*{DATA AVAILABILITY}
The data for main figures in this work is available at \url{https://github.com/Jiqiang-Zhang/Minority_Game/tree/main/DATA}
\section*{CODE AVAILABILITY}
The code for generating Fig.\ref{fig:volatility} (a) in this work is available at \url{https://github.com/Jiqiang-Zhang/Minority_Game/tree/main/Code}

\appendix
\renewcommand{\thefigure}{\Alph{section}.\arabic{figure}} 
\section{The analysis for volatility}\label{sec:app_volatility}
In Fig.~\ref{fig:timeseries_Pearson} (a-c), it can be observed that $\bar{f}_{1}$ in the population always approaches $1/2$, while both $\bar{f}_{c_1}$ and $\bar{f}_{q_1}$ in subpopulations deviate from $1/2$. Here, we first focus on the factors that influence the deviations $\Delta\bar{f}_{c_1}$ and $\Delta\bar{f}_{q_1}$. The definition of $\bar{f}_{c_1(q_1)}$ can be reformulated as a new form 
\begin{align}
	\bar{f}_{c_1(q_1)} &:=\frac{\sum\limits_{t=t_0}^{T}f_{c_{1}(q_1)}(\tau)}{T-t_0} \nonumber \\
	&= \frac{1}{2} +  \frac{\sum\limits_{t=t_0}^{T}\Delta f_{c_{1}(q_1)}(\tau)}{T-t_0} \nonumber\\
	&= \frac{1}{2}+\Delta \bar{f}_{c_1(q_1)}
\end{align}
According to relation between $\bar{f}_1$ and $\bar{f}_{c_1(q_1)}$, and $\bar{f}_1\approx 1/2$, we have  
\begin{align}
\bar{f}_1 &= \frac{1}{T - t_0} \sum_{\tau = t_0}^{T} \left( f_cf_{c_1}(\tau) + f_qf_{q_1}(\tau) \right) \nonumber \\
	&= \frac{f_c+f_q}{2}+\frac{\sum\limits_{\tau = t_0}^{T} \left[ \left(f_c\Delta f_{c_1}(\tau) + f_q\Delta f_{q_1}(\tau) \right) \right]}{T - t_0} \nonumber\\
	& = \frac{f_c+f_q}{2} + f_{c}\Delta \bar{f}_{c_1} + f_{q}\Delta \bar{f}_{q_1} \approx \frac{1}{2}. 
\end{align}
Under the normalization $f_{c} + f_{q} = 1$, one learns that 
\begin{align}
	f_{c}\Delta\bar{f}_{c_1}+f_{q}\Delta\bar{f}_{q_1} \approx 0.
\end{align}
This indicates that there exists a strong negative correlation between $\Delta\bar{f}_{c_1}$ and $\Delta\bar{f}_{q_1}$, and the absolute value $|\Delta\bar{f}_{c_1(q_1)}|$ is also negatively correlated with the fraction of its subpopulation within the entire population.

In the following, our focus is on the relation between the volatility $\psi$ and C-volatility $\psi_{c}$, Q-volatility $\psi_{q}$. First,
we rewrite the definition of $\psi_{c(q)}$ of Eq.~\eqref{eq:sub_volatilities} as follows  
\begin{align}\label{eq:app_psicq}
	\psi_{c(q)} 
	& = \frac{\sum\limits_{\tau = t_0}^{T}\left(N_{c_1(q_1)}(\tau) - \overline{N}_{c_1(q_1)}\right)^2}{|\mathcal{N}_{c(q)}|(T-t_0)}\nonumber\\
	& =  \frac{|\mathcal{N}|f_{c(q)}}{T-t_0}\sum\limits_{\tau=t_0}^{T}(f_{c_1(q_1)}(\tau)-\bar{f}_{c_1(q_1)})^2.
\end{align}
Similarly, $\psi$ in Eq.~\eqref{eq:volatility} and $r$ in Eq.~\eqref{eq:pearson_corraltion} can be reformulated as 
\begin{align}\label{eq:app_psi}
	\psi &= \frac{\sum\limits_{\tau = t_0}^{T}\left(N_{1}(\tau) - C_1\right)^2}{|\mathcal{N}|(T-t_0)} = \frac{\sum\limits_{\tau = t_0}^{T}\left(N_{1}(\tau) - \overline{N}_{1}\right)^2}{|\mathcal{N}|(T-t_0)} \nonumber\\
	& = \displaystyle{\frac{|\mathcal{N}|}{(T-t_0)|} \sum_{\tau=t_0}^{T}\left[f_c(f_{c_1}(\tau)-\bar{f}_{c_1})+f_q(f_{q_1}(\tau)-\bar{f}_{q_1})\right]^2} \nonumber \\
	& = f_{c}\psi_{c}+f_{q}\psi_{q} + 2f_cf_q\frac{|\mathcal{N}|}{T-t_0}\times\nonumber\\
	&\quad\sum_{\tau = t_0}^{T}\left(f_{c_1}(\tau)-\bar{f}_{c_1}\right)\left(f_{q_1}(\tau)-\bar{f}_{q_1}\right).
\end{align}
and  
\begin{align}
	r &= \frac{\sum\limits_{\tau=t_0}^{T}\left(N_{c_1}(\tau)-\overline{N}_{c_1}\right)\left(N_{q_1}(\tau)-\overline{N}_{q_1}\right)}
	{\sqrt{{\sum\limits_{\tau = t_0}^{T}(N_{c_1}(\tau)-\overline{N}_{c_1})^2}}{\sqrt{{\sum\limits_{\tau = t_0}^{T}(N_{q_1}(\tau)-\overline{N}_{q_1})^2}}}}, \nonumber\\
	& = \frac{|\mathcal{N}|}{(T-t_0)}\frac{\sqrt{f_cf_q}}{\sqrt{\psi_c\psi_q}}\sum_{\tau=t_0}^{T}\left[\left(f_{c_1}(\tau)-\bar{f}_{c_1}\right)\left(f_{q_1}(\tau)-\bar{f}_{q_1}\right)\right].
	\label{eq:app_pearson_corraltion}
\end{align}
Based on Eqs.~(\ref{eq:app_psicq}-\ref{eq:app_pearson_corraltion}), we can get the relation between $\psi$ and  $\psi_c$, $\psi_q$ that is 
\begin{align}\label{app_eq:psi_psic_psiq}
	\psi &=  f_{c}\psi_{c}+f_{q}\psi_{q} + 2r\sqrt{f_{c}\psi_{c}}\sqrt{f_q\psi_{q}}.
\end{align}
The relation also provides another semi-analytical method to
calculate the Pearson coefficient $r$ according to the given $f_c$, and simulated $\psi$, $\psi_c$, $\psi_q$. The result is presented in Fig. \ref{fig:timeseries_Pearson}(d), which is consistent with the full simulation.
\section{The analysis of synchronization and anti-synchronization of clusters in Q-population}\label{sec:app_syn_antisyn}
Analogy with Eq.~\eqref{app_eq:psi_psic_psiq} we can get the Q-volatility that is
\begin{align}
	\psi_{q} = f_{qq}\psi_{qq} + f_{qc}\psi_{qc} + 2r^{\prime}\sqrt{f_{qq}\psi_{qq}}\sqrt{f_{qc}\psi_{qc}}
\end{align}
with
\begin{subequations}
	\begin{empheq}[left=\empheqlbrace]{align}
		&f_{qq} := \frac{|\mathcal{C}_{q}^{\text{\RNum{1}}}|+|\mathcal{C}_{q}^{\text{\RNum{2}}}|}{|\mathcal{N}_{q}|},\quad f_{qc} := \frac{|\mathcal{C}_{q}^{\text{\RNum{3}}}|}{|\mathcal{N}_{q}|}; \\
		& N_{q_1}^{\text{\RNum{1}(\RNum{2},\RNum{3})}}(\tau) := \sum_{i\in\mathcal{C}^{\text{\RNum{1}(\RNum{2},\RNum{3})}}_q}a^{i}(\tau), \\ 
		&\overline{N}_{q_1}^{\text{\RNum{1}(\RNum{2},\RNum{3})}} := \frac{\sum\limits_{\tau = t_0}^{T}N_{q_1}^{\text{\RNum{1}(\RNum{2},\RNum{3})}}(\tau)}{T-t_0}; \\
		&\psi_{qq} := \frac{\sum\limits_{\tau = t_0}^{T}\left(N_{q_1}^{\text{\RNum{1}}}(\tau)+N_{q_1}^{\text{\RNum{2}}}(\tau) -
			\overline{N}_{q_1}^{\text{\RNum{1}}}-\overline{N}_{q_1}^{\text{\RNum{2}}}\right)^2}{\left(|\mathcal{C}_{q}^{\text{\RNum{1}}}|+|\mathcal{C}_{q}^{\text{\RNum{2}}}|\right)(T-t_0)},  \\
		& \psi_{qc} := \frac{\sum\limits_{\tau = t_0}^{T}\left(N_{q_1}^{\text{\RNum{3}}}(\tau)-\overline{N}_{q_1}^{\text{\RNum{3}}}\right)^2}{|\mathcal{C}_{q}^{\text{\RNum{3}}}|\left(T-t_0\right)}; 
	\end{empheq}
	\label{app_eqs:fqq_psiqq_pisqc}
\end{subequations}
and
\begin{align}\label{app_eq:Pearson_IS-clusters_ES-cluster}
	r^{\prime} &:= \sum\limits_{\tau=t_0}^{T}\left[\frac{N^{\text{\RNum{1}}}_{q_1}(\tau) + N^{\text{\RNum{2}}}_{q_1}(\tau) -\overline{N}^{\text{\RNum{1}}}_{q_1} - \overline{N}^{\text{\RNum{2}}}_{q_1}}{\sqrt{{\sum\limits_{\tau = t_0}^{T}(N^{\text{\RNum{1}}}_{q_1}(\tau) + N^{\text{\RNum{2}}}_{q_1}(\tau) -\overline{N}^{\text{\RNum{1}}}_{q_1} - \overline{N}^{\text{\RNum{2}}}_{q_1})^2}}}\times\right. \nonumber \\
	&\left.\qquad\qquad \frac{N^{\text{\RNum{3}}}_{q_1}(\tau)-\overline{N}^{\text{\RNum{3}}}_{q_1}}{{\sqrt{{\sum\limits_{\tau = t_0}^{T}(N^{\text{\RNum{3}}}_{q_1}(\tau)-\overline{N}^{\text{\RNum{3}}}_{q_1})^2}}}}\right].
\end{align}
Here, $N_{q_1}^{\text{\RNum{1}(\RNum{2},\RNum{3})}}$ is the number of agents entering resource $1$
in $\mathcal{C}^{\text{\RNum{1}(\RNum{2},\RNum{3})}}_q$. And, $r^{\prime}$ is the Pearson correlation coefficient between the IS-clusters $\mathcal{C}_{q}^{\text{\RNum{1}}}$ and $\mathcal{C}_{q}^{\text{\RNum{2}}}$, and the ES-cluster $\mathcal{C}_{q}^{\text{\RNum{3}}}$.
In Eq.~\eqref{app_eq:psi_psic_psiq}, our focus is on $\psi_{qq}$,which
denotes the volatility of the IS-clusters $\mathcal{C}^{\text{\RNum{1}}}_q$ and
$\mathcal{C}^{\text{\RNum{2}}}_q$. Considering that the cluster division is accomplished via a clustering algorithm, we assume that both IS-clusters $\mathcal{C}^{\text{\RNum{1}}}_q$ and $\mathcal{C}^{\text{\RNum{2}}}_q$ possess perfect intra-synchronization. That is to say, the agents in the same IS-cluster always take the same action at each step.

According to the assumption, we have
\begin{align}\label{app_eq:psi1_q1}
	\mathbb{E}(N_{q_1}^{\text{\RNum{1}}}(\tau)) = \mathbb{E}(\sum\limits_{i\in\mathcal{C}_{q}^{\text{\RNum{1}}}} a^{i}(\tau)) \in\set{|\mathcal{C}^{\text{\RNum{1}}}_q|(1-\frac{\epsilon}{2}),|\mathcal{C}^{\text{\RNum{1}}}_q|\frac{\epsilon}{2}}.
\end{align}
Based on the synchronization factor, the expected
number of agents in $\mathcal{C}^{\text{\RNum{2}}}_q$ that enter resource $1$ is
\begin{align}\label{app_eq:psi2_q1}
	N^{\text{\RNum{2}}}_{q_1}(\tau) &=\frac{|\mathcal{C}^{\text{\RNum{2}}}_{q}|}{|\mathcal{C}^{\text{\RNum{1}}}_{q}|}\left[\left(1-\langle\sigma^{\text{\RNum{1},\RNum{2}}}_q\rangle\right)\left(|\mathcal{C}^{\text{\RNum{1}}}_q|-N^{\text{\RNum{1}}}_{q_1}(\tau)\right)+\langle\sigma^{\text{\RNum{1},\RNum{2}}}_q\rangle N^{\text{\RNum{1}}}_{q_1}(\tau)\right] \nonumber\\
	& = |\mathcal{C}^{\text{\RNum{2}}}_{q}|\left(1-\langle\sigma^{\text{\RNum{1},\RNum{2}}}_q\rangle\right) + \frac{|\mathcal{C}^{\text{\RNum{2}}}_q|}{|\mathcal{C}^{\text{\RNum{1}}}_{q}|}\left(2\langle\sigma^{\text{\RNum{1},\RNum{2}}}_q\rangle-1\right) N^{\text{\RNum{1}}}_{q_1}(\tau).
\end{align}
Here, $\langle\sigma^{\text{\RNum{1},\RNum{2}}}_q\rangle$ represents the average synchronization factor be-
tween any agents that respectively belong to $\mathcal{C}^{\text{\RNum{1}}}_{q}$ and $\mathcal{C}^{\text{\RNum{2}}}_{q}$.
Then, the total volatility of $\mathcal{C}^{\text{\RNum{1}}}_{q}$ and $\mathcal{C}^{\text{\RNum{2}}}_{q}$ is that
\begin{align}\label{app_eq:psi_qq}
	\psi_{qq} & =  \frac{\sum\limits_{\tau = t_0}^{T}\left(N_{q_1}^{\text{\RNum{1}}}(\tau)+ N_{q_1}^{\text{\RNum{2}}}(\tau) -
		\overline{N}_{q_1}^{\text{\RNum{1}}}-\overline{N}_{q_1}^{\text{\RNum{2}}}\right)^2}{(|\mathcal{C}_{q}^{\text{\RNum{1}}}|+|\mathcal{C}_{q}^{\text{\RNum{2}}}|)(T-t_0)} \nonumber\\
		& = \frac{\left[1+\frac{|\mathcal{C}^{\text{\RNum{2}}}_{q}|}{|\mathcal{C}^{\text{\RNum{1}}}_{q}|}2(\langle\sigma^{\text{\RNum{1},\RNum{2}}}_q\rangle-1)\right]^2\cdot \sum\limits_{\tau=t_0}^{T}\left[N_{q_1}^{\text{\RNum{1}}}(\tau)-\overline{N}_{q_1}^{\text{\RNum{1}}}\right]^2}{(T-t_0)(|\mathcal{C}^{\text{\RNum{1}}}_{q}|+|\mathcal{C}^{\text{\RNum{2}}}_{q}|)}.
\end{align}

For the optimal intra-synergy of the Q-subpopulation, two conditions need to be met: (\Rnum{1}) the volatility $\psi_{qq} \approx 0$ and
(\Rnum{2}) the expected number of agents entering the resource $1$ in total IS-clusters is $N^{\text{\RNum{1}}}_{q_1}(\tau) + N^{\text{\RNum{2}}}_{q_1}(\tau) = (|\mathcal{C}^{\text{\RNum{1}}}_{q}|+|\mathcal{C}^{\text{\RNum{2}}}_{q}|)/2$. According to Eq.~\ref{app_eq:psi_qq}, one learns that there are two ways to meet Condition (\Rnum{1}) as follows
\begin{subequations}\label{app_eq:conditions}
	\begin{empheq}[left=\empheqlbrace]{align}
		  &N^{\text{\RNum{1}}}_{q_1}(\tau) \approx \overline{N}_{q_1}^{\text{\RNum{1}}} \label{app_eq:condition1}, \\
		  &|\mathcal{C}^{\text{\RNum{1}}}_q|  \approx |\mathcal{C}^{\text{\RNum{2}}}_q|(1-2\langle\sigma^{\text{\RNum{1},\RNum{2}}}_q\rangle).
		  \label{app_eq:condition2}
	\end{empheq}
\end{subequations}
Equation~\eqref{app_eq:condition1} suggests that the Condition (\Rnum{1}) can be fulfilled by means of nearly static intra-synchronization IS-clusters $\mathcal{C}^{\text{\RNum{1}}}_q$ and $\mathcal{C}^{\text{\RNum{2}}}_q$, which is consistent with the results in Fig.~\ref{fig:timeseries_action}(a-b).
This indicates that, because of the static intra-synchronization in IS-clusters $\mathcal{C}^{\text{\RNum{1}}}_q$ and $\mathcal{C}^{\text{\RNum{2}}}_q$, the average synchronization factor $\langle\sigma^{\text{\RNum{1},\RNum{2}}}_q\rangle_q$ gets close to either 
$1$ or $0$. Nevertheless, $\mathcal{C}^{\text{\RNum{1}}}_q$ and $\mathcal{C}^{\text{\RNum{2}}}_q$ are two distinct clusters rather than a single merged one. Consequently, $\langle\sigma^{\text{\RNum{1},\RNum{2}}}_q\rangle \approx 0$. This result implies that IS-clusters $\mathcal{C}^{\text{\RNum{1}}}_q$ and $\mathcal{C}^{\text{\RNum{2}}}_q$ exhibit inter-anti-synchronization. Moreover, the two methods described in Eq.~\eqref{app_eq:conditions} can work together to further reduce volatility $\psi_{qq}$. On the basis of Eq.~\eqref{app_eq:condition2}, we can get $|\mathcal{C}^{\text{\RNum{1}}}_q| = |\mathcal{C}^{\text{\RNum{2}}}_q|$ under $\langle\sigma^{\text{\RNum{1},\RNum{2}}}_q\rangle \approx 0$, i.e., the sizes of the intra-synergic clusters are approximately the same. 

However, with the increase of $f_c$, Fig.~\ref{fig:timeseries_action} shows the synchronization gradually changes from static to dynamic. Then, $N^{\text{\RNum{1}}}_{q_1}(\tau)\not\approx\overline{N}_{q_1}^{\text{\RNum{1}}}$. This means Eq.~\eqref{app_eq:condition2} is the only way
to meet the Condition (\Rnum{1}). In the case, Fig.~\ref{fig:timeseries_Pearson} demonstrates the anti-synchronicity between $\mathcal{C}^{\text{\RNum{1}}}_q$ and $\mathcal{C}^{\text{\RNum{2}}}_q$ also increase from $0$,
i.e., $0 < \langle\sigma^{\text{\RNum{1},\RNum{2}}}_q\rangle < 0.5$. Then, based on Eq.~\eqref{app_eq:condition2}, we find that the sizes of IS-clusters $\mathcal{C}^{\text{\RNum{1}}}_q$ and $\mathcal{C}^{\text{\RNum{2}}}_q$ cannot remain equal. 

For the Condition (\Rnum{2}), after substituting Eq.~\eqref{app_eq:psi2_q1} into it, we can get a relation that is
\begin{align}
	N_{q_1}^{\text{\RNum{1}}}(\tau) - \frac{|\mathcal{C}^{\text{\RNum{1}}}_{q}|}{2} = \langle\sigma^{\text{\RNum{1},\RNum{2}}}_q\rangle  |\mathcal{C}^{\text{\RNum{2}}}_q| - \frac{|\mathcal{C}^{\text{\RNum{2}}}_{q}|}{|\mathcal{C}^{\text{\RNum{1}}}_{q}|}\left(\langle\sigma^{\text{\RNum{1},\RNum{2}}}_q\rangle-\frac{1}{2}\right)N_{q_1}^{\text{\RNum{1}}}(\tau). \nonumber
\end{align}
Nevertheless, the relation holds true as long as Eq.~\eqref{app_eq:condition2} is met. This implies that if the Condition (\Rnum{1}) is satisfied via Eq.~\eqref{app_eq:condition2}, then the Condition (\Rnum{2}) is automatically fulfilled.

To summarize, the previously mentioned analysis indicates that there exist two ways to reach the optimal intra-synergy in the Q-subpopulation by means of the IS-clusters $\mathcal{C}^{\text{\RNum{1}}}_q$ and $\mathcal{C}^{\text{\RNum{2}}}_q$. One is that both clusters are statically intra-synchronized, have the same size, and approximately show inter-anti-synchronization with each other. Another is that both IS-clusters are dynamically intra-synchronized, but they differ in size and present weak inter-anti-synchronization. Figure~\ref{fig:timeseries_action} shows the former
occurs when $f_{c}$ is low. In contrast, the latter phenomenon occurs when $f_{c}$ is high yet still below the transition point $f_{c}^{*}$.

\section{The result of K-means clustering analysis for C-subpopulation}
\begin{figure}[htbp!]
	\centering
	\includegraphics[width=\linewidth]{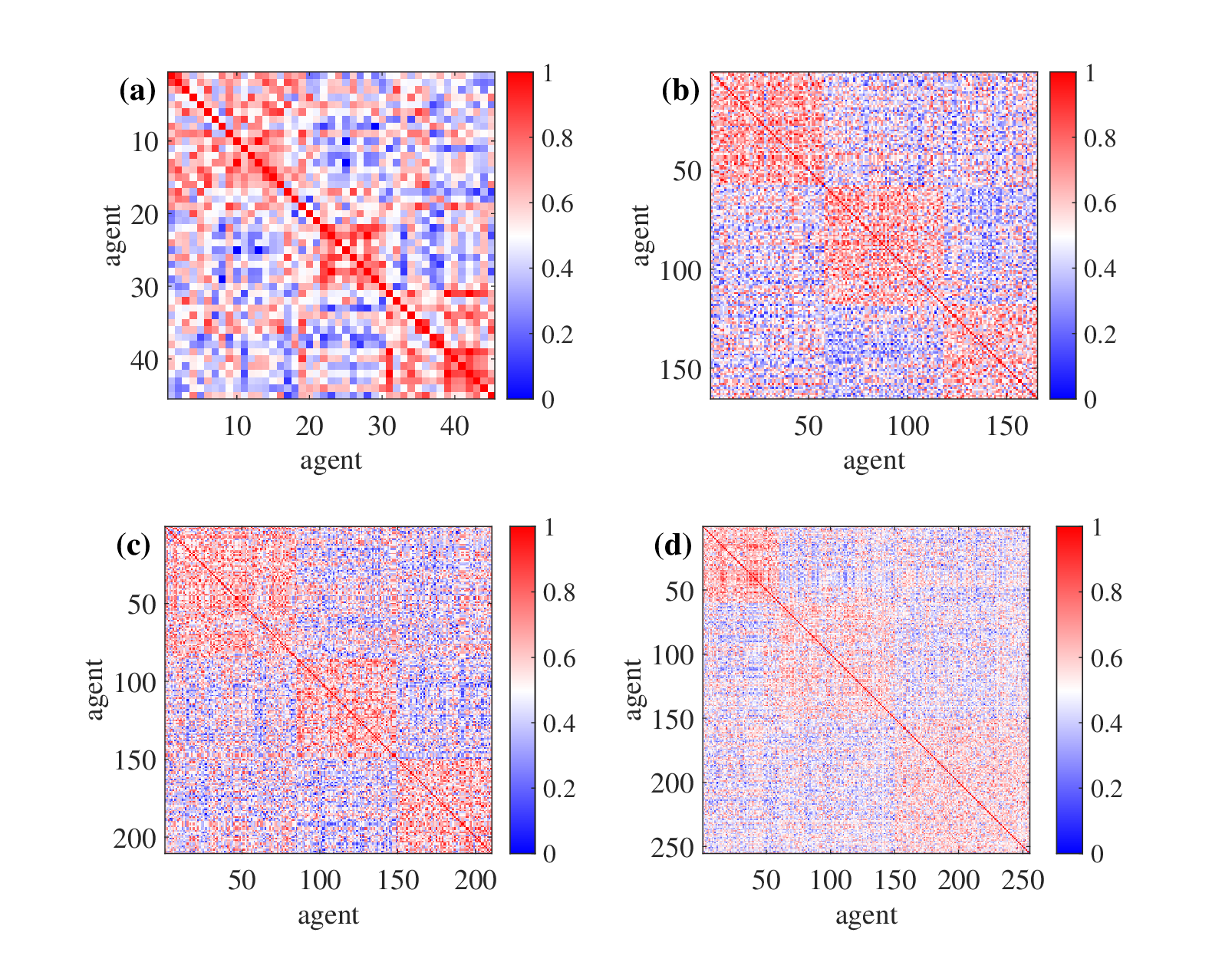}
	\caption{ (Color online) {\bf K-means clustering analysis for C-subpopulation under the synchronization.} Panels (a-d) show the K-means clustering analysis results based on the synchronization between any pair of agents $i$ and $j$ within C-subpopulation. In the results, the setup of number of cluster is $K = 3$. In (a-d), the fraction of C-population is $f_{c} = 0.15$, $0.55$, $0.7$ and $0.85$. 
	The learning parameters $(\alpha, \gamma, \epsilon)$, memory length $m$, and system size $|\mathcal{N}|$ are adopted by default as in Fig.~\ref{fig:volatility}.
	}\label{fig:C_clustering}
\end{figure}

In here, as a comparison, we also take an examination of the synchronization between any two agents $i$ and $j$ within the C-subpopulation $\mathcal{N}_{c}$. Based on the action time series, 
the synchronization between $i$ and $j$ is defined as 
\begin{align}\label{eq:synchronization_factor2}
	\sigma_{c}^{i,j} &:= 1 - \bar{d}_{H_{c}}(\bm{a}^i,\bm{a}^{j}) \nonumber\\
	&= 1 - \frac{\sum\limits_{\tau=t_0}^{T} |a^{i}(\tau)-a^{j}(\tau)|}{T-t_0}, 
\end{align}
where $\bar{d}_{H_{c}}(\bm{a}^i,\bm{a}^{j})$ is the average Hamming distance between the time series $\bm{a}^i$ and $\bm{a}^{j}$ for $i$ and $j$ that both belong to $\mathcal{N}_{c}$. Then, we also perform the K-means clustering analysis on the matrix $\bm{\sigma}_{c}$ of $\mathcal{N}_{c}$ and set the number of clusters as $K = 3$ for comparison. Similarly, the clusters obtained from the C-population are labeled as $\mathcal{C}_{c}^{\text{\RNum{1}}}$, $\mathcal{C}_{c}^{\text{\RNum{2}}}$, $\mathcal{C}_{c}^{\text{\RNum{3}}}$. The results of K-means clustering analysis for C-subpopulation are shown in Fig.~\ref{fig:C_clustering}.

Given that the state-action mapping of each method in the set is randomly generated, it is improbable that two agents within the C-subpopulation synchronize their behaviors in time series. The absence of separate clusters in the synchronous cluster analysis further confirms this, as shown in Fig.~\ref{fig:C_clustering}. 

\section{Mathematical Notation Descriptions}\label{sec:notations}    
Here, we present the descriptions of mathematical notations used in simulation and analysis, along with the locations of the definitions for the corresponding notations in Table~\ref{tab:notation2}.

\begin{table*}[htbp!]
	\caption{\label{tab:notation2}%
		{\bf The descriptions for mathematical notations in simulation and analysis}}
	\begin{ruledtabular}
		\begin{tabular}{p{3cm}p{10cm}p{4cm}}
			Symbol&  Description  & Defined by or in\\
			\colrule
			$\mathcal{N}$ & The set of populations composed of all agents & Sec.~\ref{sec:model}\\
			$\mathcal{N}_c/\mathcal{N}_q$ & The set of subpopulation composed of agents using classical-policy/Q-policy & Sec.~\ref{sec:model}\\ 
			$\psi/\psi_c/\psi_q$    &   The volatility of population/C-subpopulation/Q-subpopulation & Eqs.~(\ref{eq:volatility}-\ref{eq:sub_volatilities})\\ 
			$f_{c}/f_{q}$ & The fraction of C-subpopulation/Q-subpopulation in the entire population & Sec.~\ref{sec:model} \\
			$N_{1}/N_{c_1}/N_{q_1}$ & The number of agents entering resource $1$ in population/C-subpopulation/subQ-population & Sec.~\ref{sec:model}\\
			$f_{1}/f_{c_1}/f_{q_1}$  & The fraction of agents entering resource $1$ in the population/C-subpopulation/Q-subpopulation & Eq.~\eqref{eq:f1_fc1_fq1}\\
			$r$ &  The Pearson coefficient between the time series of $N_{c_1}$ and $N_{q_1}$ & Eq.~\eqref{eq:pearson_corraltion}\\
			$\sigma_{c}^{i,j}/\sigma^{i,j}_{q}$ &  The synchronization between two agents $i$ and $j$ within C-subpopulation/Q-subpopulation & Eq.~\eqref{eq:synchronization_factor} \\
			$\bm{\sigma}_{c}/\bm{\sigma}_{q}$ & The synchronization matrix composed of $\sigma_{c}^{i,j}/\sigma^{i,j}_{q}$ & Sec.~\ref{subsubsec:Intra_synergy} \\
			$\mathcal{C}^{\text{\RNum{1}(\RNum{2},\RNum{3})}}_{c}/\mathcal{C}^{\text{\RNum{1}(\RNum{2},\RNum{3})}}_{c}$ & The cluster sets obtained by dividing the Q-subpopulation/C-population  &  Sec.~\ref{subsubsec:syn_antisyn}\\
			$f_{qq}$ & The fraction of agents belonging to the IS-clusters $\mathcal{C}^{\text{\RNum{1}}}_{q}$ and $\mathcal{C}^{\text{\RNum{2}}}_{q}$ in Q-subpopulation &  Eq.~\eqref{app_eqs:fqq_psiqq_pisqc}		\\
			$f_{qc}$  & The fraction of agents belonging to the ES-cluster $\mathcal{C}^{\text{\RNum{3}}}_{q}$ in Q-subpopulation & Eq.~\eqref{app_eqs:fqq_psiqq_pisqc}\\
			$\psi_{qq}$ & The volatility of the IS-clusters $\mathcal{C}^{\text{\RNum{1}}}_{q}$ and $\mathcal{C}^{\text{\RNum{2}}}_{q}$ & Eq.~\eqref{app_eqs:fqq_psiqq_pisqc} \\
			$\psi_{qc}$ & The volatility of the ES-cluster $\mathcal{C}^{\text{\RNum{3}}}_{q}$ & Eq.~\eqref{app_eqs:fqq_psiqq_pisqc} \\
			$N_{q_1}^{\text{\RNum{1}}}/N_{q_1}^{\text{\RNum{2}}}/N_{q_1}^{\text{\RNum{3}}}$ & The number of agents entering resource in cluster $\mathcal{C}_{q}^{\text{\RNum{1}}}$/$\mathcal{C}_{q}^{\text{\RNum{2}}}$/$\mathcal{C}_{q}^{\text{\RNum{3}}}$               & Eq.~\eqref{app_eqs:fqq_psiqq_pisqc} \\
			$r^{\prime}$ & The Pearson coefficient between the time series of $N_{q_1}^{\text{\RNum{1}}}+N_{q_1}^{\text{\RNum{2}}}$ and $N_{q_1}^{\text{\RNum{3}}}$ & Eq.~\eqref{app_eq:Pearson_IS-clusters_ES-cluster} \\
			$\langle\sigma^{\text{\RNum{1},\RNum{2}}}_q\rangle$ &  The average synchronization be-
			tween any agents that respectively belong to $\mathcal{C}^{\text{\RNum{1}}}_{q}$ and $\mathcal{C}^{\text{\RNum{2}}}_{q}$  & Sec.~\ref{sec:app_syn_antisyn} \\
			$p(s_\mu)$ &  The prabability of state $s_{\mu}$ & Eq.~\eqref{eq:p_state} \\
			$p(r_{w} = 1|s_\mu)$ &  The conditional winning probability of resource $1$ given state $s_\mu$ at $\tau$ & Eq.~\eqref{eq:p_condition} \\
			$\lambda_{\hat{\bm{a}}}$ & The growth rate of score of strategy $\hat{\bm{a}}$ & Eq.~\eqref{eq:growth_rate} \\
			$\phi$           & The ratio of frozen agents within C-population    &  Eq.~\eqref{eq:frozen_ratio}  \\
			$F(n)$          & The cumulative distribution function of the number of switches of the optimal strategy for agents in C-subpopulation       & Sec.~\ref{sec:analysis_csubpopulation}
		\end{tabular}
	\end{ruledtabular}
\end{table*}

\bibliography{document}

\end{document}